\documentclass[review]{elsarticle}

\usepackage{lineno,hyperref}
\usepackage{subfigure}
\usepackage{amsmath, amssymb}
\usepackage{amsbsy}
\usepackage{mathptmx}
\usepackage{bm}
\usepackage{makecell}
\modulolinenumbers[5]

\journal{XXX}

\begin{document}

\begin{frontmatter}

\title{An extending strategy based on  TENO framework  for      hyperbolic conservation laws}

\author[mymainaddress,myforthaddress]{Fan Zhang}
\ead{a04051127@mail.dlut.edu.cn,zhangfan3@sysu.edu.cn}

\author[mysecondaryaddress]{Jun Liu}
\author[mythirdaddress]{Huaibao Zhang}

\author[mymainaddress]{Chunguang Xu\corref{mycorrespondingauthor}}
\ead{xuchg5@sysu.edu.cn}
\cortext[mycorrespondingauthor]{Corresponding author.}

\address[mymainaddress]{School of Aeronautics and Astronautics, Sun Yat-sen University, Guangzhou, China }
\address[mysecondaryaddress]{School of Aeronautics and Astronautics, Dalian University of Technology, Dalian, China}

\address[mythirdaddress]{School of Physics, Sun Yat-sen University, Guangzhou, China }
\address[myforthaddress]{State Key Laboratory of Aerodynamics, China Aerodynamics Research and Development Center, Mianyang, China }
\begin{abstract}
Recently, the targeted ENO (TENO) schemes  give a novel framework to keep optimal high-order spatial reconstruction  wherever discontinuity is deemed to be vanished, including at smooth critical points, and to avoid oscillations by completely removing stencils crossing discontinuities.
Moreover, the smoothness measurement of TENO schemes is in fact acting as shock-detectors, which are capable for distinguishing  discontinuities and smooth critical points.
Following the idea of a recent improvement, i.e. TENO-NA, the shock-detection and stencil-selection are completely separated in this work.
Higher-order polynomials using neighbouring points of the standard five-point TENO scheme are applied for achieving higher-order accuracy without significantly increasing computation  cost, by exploring the neighbouring smoothness measurements. In this work, seventh-order spatial accuracy is achieved, and the computational complexity is similar to that of the five-point TENO scheme. Especially, the presented method introduces new flexibility in constructing high-order numerical methods.
Numerical results are given to show the shock-capturing and wave-resolving capabilities.

\end{abstract}


\end{frontmatter}

\section{Introduction} \label{sec:Intro}

The spatial solutions of flow field containing discontinuities are extremely challenging, because of the nonlinear essential of the governing equations. Especially, achieving  high-order accuracy on smooth flow field including critical points and eliminating oscillation near discontinuities are two crucial issues that are attracting attentions, due to their necessities as well as their difficulties.
Without stressing the details here, we only need to recall the fact that a high-order interpolation will produce spurious oscillation near discontinuities, leading to troublesome problems such as  positivity issues and erroneous solutions.

In order to solve those problems, various methodologies have been developed. One of the most popular high-order numerical methods for solving hyperbolic conservation laws is the WENO scheme \cite{Liu1994} which utilises all the candidate stencils of ENO scheme \cite{Harten1987} by using  a nonlinear weighting strategy.  WENO-type schemes approximately recover the background linear schemes if discontinuities are deemed to be vanished, according to the local and   global smoothness indicators  distinguishing smooth field and discontinuities, leading to high-order of accuracy in smooth field. At the meantime, oscillating stencils are approximately eliminated by  decreasing the corresponding nonlinear weights, since they are likely to be crossing discontinuities. Numerous contributions have been presented to improve the performance of WENO schemes, including the accuracy in smooth field and the shock-capturing capability. There are several popular WENO schemes, such as WENO-JS scheme \cite{Jiang1996}, WENO-M scheme \cite{Henrick2005} and WENO-Z scheme \cite{Borges2008}, to name a few.
Due to their shock-capturing capability and the high-resolution property, WENO schemes have been applied in various application areas, including LES
(Large eddy simulation) \cite{Kokkinakis2015,Ritos2018a,Ritos2018} and DNS (Direct numerical simulation) \cite{Li2010,Tong2017}, etc.

Recently, Targeted ENO (TENO) schemes have been invented by Fu et al. \cite{Fu2016}. Instead of reducing the contribution of oscillatory stencils
by the nonlinear weighting strategy of WENO schemes, TENO schemes completely remove the oscillatory stencils in the final reconstruction, and fully recover the background linear scheme if all the candidate stencils are deemed to be smooth. So far, various numerical results including turbulence flows \cite{Fu2017}, multi-phase flows \cite{Haimovich2017} and detonations \cite{zhangcicp_2018}, have proved that TENO schemes are accurate and robust. Most importantly, TENO schemes, in fact,  give a new way to fully discard the weighting procedure  \cite{Fu2018}, facilitating the utilization of the candidate stencils and the development of higher-order schemes.

Usually, in order to increase the order of accuracy of a specific numerical scheme, more degree-of-freedoms (DOFs) are necessary for constructing  higher-order polynomials. For finite difference schemes such as WENO or TENO, more DOFs are usually given by extending the candidate stencil(s). For example, instead of using three three-point stencils in a five-point WENO scheme, the seven-point WENO scheme uses four four-point stencils \cite{Balsara2000}. At the meantime, more complicate smoothness indicators are required for measuring oscillation and calculating the nonlinear weights, leading to higher computation cost. Since TENO schemes are developed to function as a discontinuity-location detector \cite{Fu2018} and the high-order polynomial can be calculated separately, simpler smoothness indicators could be used for implementing higher-order spatial approximation, reducing computation cost.

Therefore, an extending strategy based on the TENO framework is given in this work. The following sections are organized as follows. In the next section, WENO and TENO schemes are briefly introduced, and especially the ENO-like stencil-selection procedure is specifically explained for introducing the following discussions. In section \ref{sec:new}, the recent idea in \cite{Fu2018} is briefly discussed. In section \ref{sec:method}, the presented method is introduced in detail.
In section \ref{sec:results}, several typical numerical cases are used to examine the performance of the presented method. Finally, the concluding remarks are given in the last section.

\section{Hyperbolic conservation law and (W/T)ENO schemes}

Aerodynamics or gas dynamics problems are usually described by a hyperbolic system, and one-dimensional scalar hyperbolic equations share the main property of hyperbolic systems.
 Therefore, without loss of generality,   WENO schemes and TENO schemes are introduced based on the one-dimensional scalar hyperbolic conservation law, and the schemes can be extended to hyperbolic system equations straightforwardly.

\subsection{The fundamental of the spatial solution}

The one-dimensional hyperbolic conservation law  is written as
\begin{equation}\label{eq:hcl}
\frac{{\partial u}}{{\partial t}} + \frac{{\partial f(u)}}{{\partial x}} =0 ,
\end{equation}

\noindent in which the  characteristic  velocity is $\frac{{\partial f(u)}}{{\partial u}}$  and   assumed  to   be  positive, without loss of generality. Here,
the spatial discretization of Eq.\eqref{eq:hcl}  is given on an  equally spaced one-dimensional mesh, leading to an ODE (ordinary differential equation) system, i.e.

\begin{equation}\label{eq:dis}
\frac{{d u_i}}{{d t}}=- \frac{{\partial f}}{{\partial x}}|_{x=x_i}, \quad i=1, \cdots, n.
\end{equation}

\noindent The partial derivatives in $x$-direction are approximated by using the finite difference formula, i.e.

\begin{equation}\label{eq:semi}
\frac{{d u_i}}{{d t}}=- \frac{1}{\Delta x}(h_{i+1/2}-h_{i-1/2}).
\end{equation}

 The flux function $h_{i\pm 1/2}$ at half points can be implicitly defined by
 \begin{equation}\label{eq:flux}
f(x)=\frac{1}{\Delta x}\int_{x-\Delta x /2}^{x+\Delta x /2} h(\xi)d\xi,
\end{equation}

 \noindent and   the
 semi-discretized form can be   written as
\begin{equation}\label{eq:app}
\frac{{d u_i}}{{d t}}\approx - \frac{1}{\Delta x}(\hat{f}_{i+1/2}-\hat{f}_{i-1/2}),
\end{equation}
\noindent where the numerical flux functions $\hat{f}_{i\pm 1/2}$ are calculated from the convex combination of $r$ candidate-stencil fluxes
\begin{equation}\label{eq:weighted}
\hat{f}_{i\pm 1/2}=\sum\limits_{k=0}^{r-1}\omega_k\hat{f}_{k,i\pm1/2}.
\end{equation}

In order to obtain a $(2r-1)$-order approximation for flux functions $\hat{f}_{i\pm 1/2}$,  the $(r-1)$-order interpolation on each candidate
stencil is given as

\begin{equation}\label{eq:r_1}
 h(x)\approx\hat{f}_k(x)=\sum\limits_{l=0}^{r-1}a_{l,k}x^l,
\end{equation}
\noindent where the coefficients $a_{l,k}$ can be calculated by substituting Eq.\eqref{eq:r_1} into Eq.\eqref{eq:flux} and solving
the resulting linear algebraic system.


 By applying   the spatial approximation of the flux function, the temporal differential term in the ODE system, i.e. Eq.\eqref{eq:dis}, can be solved by using the  third-order strongly stable Runge-Kutta method \cite{Gottlieb2001}, but the detailed formula is omitted for simplicity.
\subsection{Canonical  WENO  schemes}

The   fifth-order WENO-JS scheme  \cite{Jiang1996} is very popular ever since it was invented, and higher-order schemes \cite{Balsara2000} were developed based on the JS weights. For the fifth-order WENO-JS scheme, of which  $r=3$, two-degree polynomial approximation of the numerical flux function can be given as
\begin{equation}
\hat{f}_k(x)=a_{0,k}+a_{1,k}x+a_{2,k}x^2, \quad k=0, 1, 2.
\end{equation}
\noindent
 Specifically,  the numerical flux functions of the candidate stencils for the approximation at grid half point $i+\frac{1}{2}$ are
\begin{equation} \label{eq:3_rd}
\begin{split}
&\hat{f}_{0,i+1/2}=\frac{1}{6}(2f_{i-2}-7f_{i-1}+11f_i),\\
&\hat{f}_{1,i+1/2}=\frac{1}{6}(-f_{i-1}+5f_{i}+2f_{i+1}),\\
&\hat{f}_{2,i+1/2}=\frac{1}{6}(2f_{i}+5f_{i+1}-f_{i+2}).
\end{split}
\end{equation}

\noindent The error of the approximation in  Eq.(\ref{eq:3_rd}) can be obtained by Taylor expansion analysis, i.e.
\begin{equation} \label{eq:error}
 \hat{f}_{k,i+1/2}=h_{k,i+1/2}+C_k\Delta x^3+O(\Delta x^4),
\end{equation}

\noindent where $C_k$ is a constant which is independent of $\Delta x$ but related to specific candidate stencils.

The weight of each numerical flux function of a candidate stencil, i.e. Eq.\eqref{eq:weighted}, is   defined  as
\begin{equation} \label{eq:JS}
\omega_k=\frac{\alpha_k}{\sum_{k=0}^{r-1}\alpha_k}, \quad \alpha_k=\frac{d_k}{(\beta_k+\epsilon)^q}.
\end{equation}
\noindent  The  optimal linear weights  $d_k$  of fifth-order WENO schemes are
 \begin{equation} \label{eq:opt}
d_0=0.1, \quad d_1= 0.6, \quad d_2=0.3,
\end{equation}
\noindent which will generate the fifth-order background linear scheme on a five-point full stencil, and $\epsilon=10^{-6}$ is the small value avoiding
zero denominator.
 It should be noticed that the small value also acts as a cutoff of  the smoothness measurement \cite{Castro2011}, and has been modified to avoid overwhelming small measurements \cite{Borges2008}.
The exponent is usually defined as $q=2$.

The nonlinear weights in Eq.\eqref{eq:JS} is essential to suppress the oscillations crossing discontinuities, ensuring  the essentially non-oscillatory property. In general, the nonlinear weights corresponding to oscillatory stencils will be decreased, approximately removing the contribution of those stencils. Therefore, it is crucial to measure the smoothness of the flow field.
The local smoothness indicator $\beta_{k,r}$ in the nonlinear weights determines the smoothness of each stencil to the final high order reconstruction, and is defined following Jiang and Shu \cite{Jiang1996}, as
\begin{equation} \label{eq:local_smooth}
\beta_{k,r}=\sum\limits_{j=1}^{r-1}\Delta x^{2j-1}\int_{x_{i-1/2}}^{x_{i+1/2}}\left(\frac{d^j}{d x^j}\hat{f}(x) \right)^2dx.
\end{equation}

\noindent Jiang and Shu \cite{Jiang1996} gave the explicit form of the local smoothness indicator $\beta_{k,r=3}$ of fifth-order schemes  in terms of the numerical flux function $f_i$, i.e.
\begin{equation} \label{eq:smooth_expl}
\begin{split}
&\beta_0=\frac{1}{4}(f_{i-2}-4f_{i-1}+3f_i)^2+\frac{13}{12}(f_{i-2}-2f_{i-1}+f_i)^2,\\
&\beta_1=\frac{1}{4}(f_{i-1}-f_{i+1})^2+\frac{13}{12}(f_{i-1}-2f_{i}+f_{i+1})^2,\\
&\beta_2=\frac{1}{4}(3f_{i}-4f_{i+1}+f_{i+2})^2+\frac{13}{12}(f_{i}-2f_{i+1}+f_{i+2})^2,
\end{split}
\end{equation}
\noindent where the subscript $r$ is omitted for simplicity. For higher-order schemes, references \cite{Balsara2000} and \cite{Gerolymos2009} provided detailed formulas.


The design of WENO-JS scheme is effective for removing the contribution of the stencil across discontinuity. However, WENO-JS scheme \cite{Jiang1996} generally degenerates to third-order accuracy near critical points \cite{Henrick2005}, where the first-order derivative vanishes. Since smooth critical points commonly exist in
practical simulations, it is expected that the   weights in Eq.\eqref{eq:opt} can be achieved in smooth region, or the nonlinear weights can  approximately converge  to the optimal weights as $\Delta x$ approaches zero.

 Henrick et al. \cite{Henrick2005} investigated the effective order of   WENO schemes
and suggested that satisfying
 \begin{equation} \label{eq:smooth2}
 \sum\limits_{k=0}^{2}(\omega_k-d_k)=O(\Delta x^6),
\end{equation}
\noindent and
 \begin{equation} \label{eq:smooth}
 \omega_k-d_k=O(\Delta x^3),
\end{equation}
\noindent is sufficient for retaining the overall fifth-order accuracy of the nonlinear reconstruction.

Borges et al. \cite{Borges2008} suggested that the contribution of the stencil containing discontinuity should be enhanced in the entire
reconstruction to improve the accuracy of WENO, without discarding the essentially non-oscillatory property,  and
further proposed a novel smoothness indicator exploiting the five-point full stencil of fifth-order schemes, i.e.
\begin{equation} \label{eq:tau5}
 \tau_5=|\beta_0-\beta_2|.
\end{equation}
\noindent The global smoothness indicator $\tau_5$  is of $O(\Delta x^5)$ and satisfies
\begin{equation}
\frac{\tau_5}{\beta_k+\epsilon}=O(\Delta x^3), \quad k=0, 1, 2.
\end{equation}
\noindent For higher-order WENO-Z schemes, references \cite{Castro2011} and \cite{Don2013} provided detailed formulas and analysis.

Then the new weighting strategy is given as
\begin{equation}
 \alpha_k^{*}=d_k\left[ 1+ \left( \frac{\tau_5}{\beta_k+\epsilon}\right)^q\right],
\end{equation}
\noindent where  $\epsilon=10^{-40}$, and then the final weights $\omega_k^{(Z)}$ of WENO-Z scheme, are calculated by replacing   $\alpha_k$ in Eq.\eqref{eq:JS} with $\alpha_k^{*}$. By defining $q=1$, WENO-Z scheme will be relatively less dissipative comparing with using larger $q$ \cite{Borges2008}.
By using this improved nonlinear weight, WENO-Z scheme shows better resolution, especially at smooth critical points.

It should be noticed that, for higher-order upwind WENO schemes, e.g. seventh-order WENO schemes \cite{Balsara2000}, the formulas of the smoothness measurement   are more complicate, but the idea is similar to the fifth-order schemes. Specifically, the candidate stencils are all of the same width in a typical seventh-order WENO scheme. Therefore, the details of seventh-order WENO schemes are not further introduced in this article.

\subsection{TENO schemes}

TENO schemes have been systematically introduced and further developed by Fu et al. \cite{Fu2016,Fu2017,Fu2018}. There are various beneficial  features of TENO schemes. However, two features are more essential for introducing the method in this article. Firstly, the   stencils of incrementally increasing width,
with which
arbitrary high-order spatial accuracy are achieved under TENO framework, are introduced. Secondly, the ENO-like stencil-selection procedure that ensures recovering the background linear schemes is a necessity needed to be introduced.

\subsubsection{Incremental-width stencils}

\begin{figure}[h!t]
\begin{center}
\subfigure[\label{fig:f:7order2}{The candidate stencils of the seventh-order TENO scheme}]{
\resizebox*{7.5cm}{!}{\includegraphics{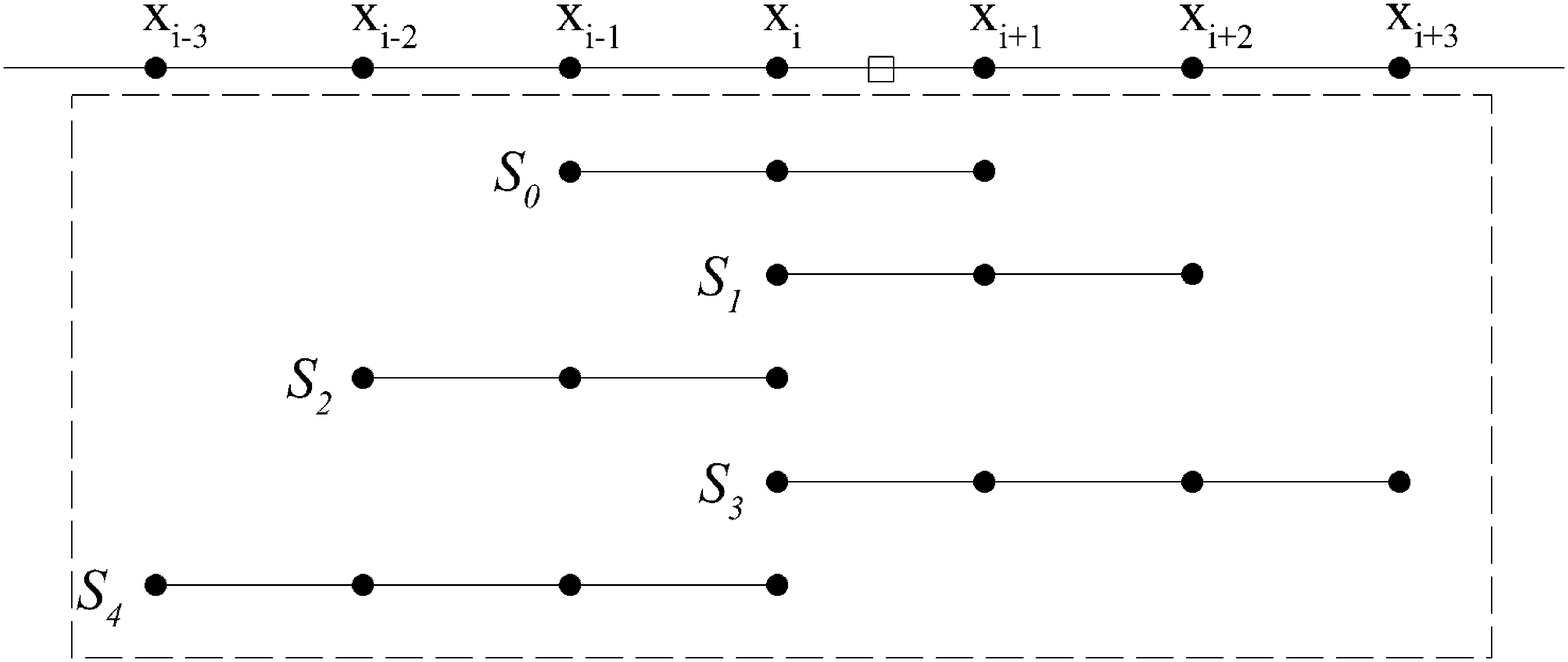}}}
\subfigure[\label{fig:f:7order1}{The candidate stencils of seventh-order WENO schemes}]{
\resizebox*{7.5cm}{!}{\includegraphics{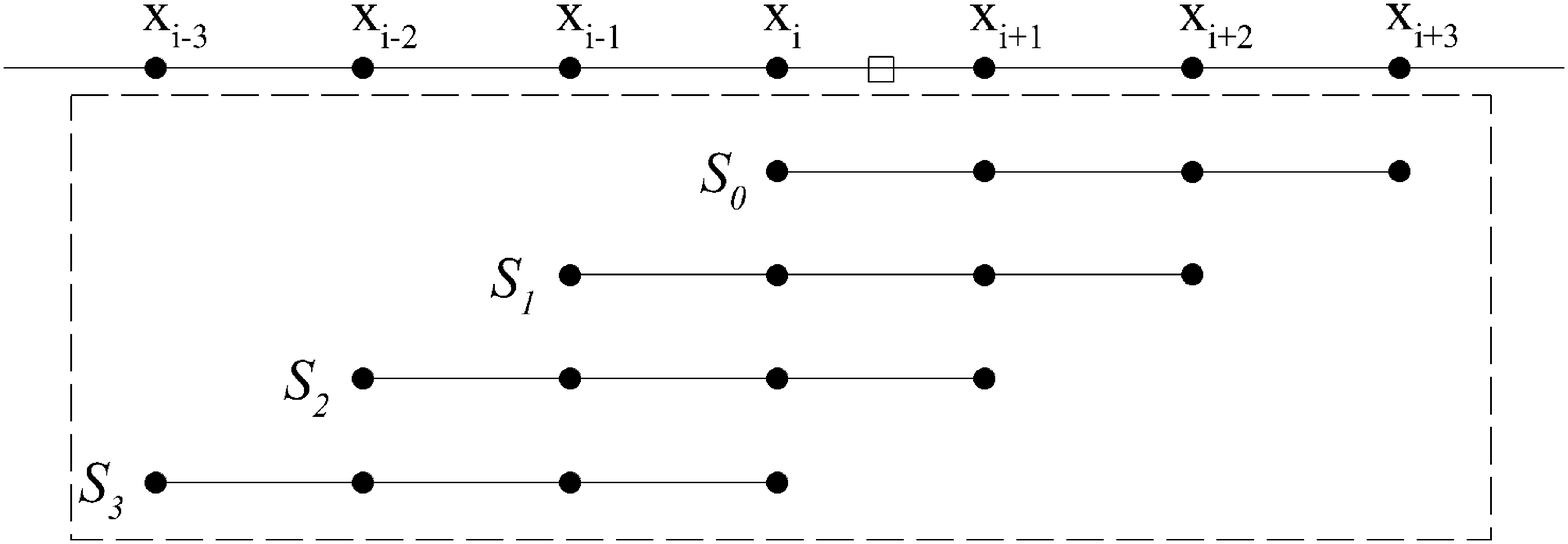}}}
\caption{\label{fig:f:7order} Schematics illustrating the candidate stencils of seventh-order WENO and TENO schemes. }
\end{center}
\end{figure}

The candidate stencils of the fifth-order TENO scheme are the same as those of WENO schemes. However, incrementally increasing width stencils are used in higher-order TENO schemes.  For attaining seventh-order spatial accuracy on a seven-point full stencil, TENO schemes use two four-point stencils, in addition to the three three-point stencils of the fifth-order TENO scheme. The stencils are schematically shown in Fig.\ref{fig:f:7order2}. The seventh-order WENO schemes use four four-point stencils, as shown in Fig.\ref{fig:f:7order1}, and in a very high order WENO scheme, the large sub-stencils enhance encounters of interacting characteristics, requiring specific treatment to avoid the nonexistence of a smooth stencil \cite{Gerolymos2009}. Therefore, TENO schemes are more suitable to deal with closely  located   shock   waves such as those in compressible turbulence problem.

The details of the seventh-order or even higher order TENO schemes, including the construction of candidate stencils and the optimal linear weights of those stencils, were introduced in \cite{Fu2016} and are not further discussed here since they are not of the major concern in this work.

\subsubsection{The ENO-like stencil-selection procedure}

Although the incremental-width candidate stencils are very eye-catching, the ENO-like stencil-selection procedure is no less important. Especially, the procedure helps TENO schemes separating the shock/discontinuity-detection and high-order spatial reconstruction, allowing significant flexibility.

Motivated by  Borges  et   al.  \cite{Borges2008} and  Hu   et   al. \cite{Hu2011}, the smoothness measurement of  TENO schemes with $K$-point full stencil is given as
 \begin{equation} \label{eq:TENO1}
\gamma_k=\left(C+\frac{\tau_K}{\beta_{k,r}+\epsilon} \right)^q, \quad k= 0, \cdots, K-3.
\end{equation}

\noindent The local smoothness indicator $\beta_{k,r}$ is the same as those in WENO schemes. The global reference smoothness indicator  $\tau_K$ was detailed in \cite{Fu2016}, and for the fifth-order TENO scheme $\tau_5$ is reused from WENO-Z scheme, as in Eq.\eqref{eq:tau5}. The small threshold also remains the value of WENO-Z scheme, i.e. $\epsilon=10^{-40}$. $C$ is set as 1, and the integer  power  $q$ is set as 6. As introduced by Fu et al. \cite{Fu2016}, larger integer power exponent $q$ and smaller $C$ are preferable for a stronger separation between resolved
and non-resolved scales, and the discontinuity-detection capability can be significantly
enhanced.

However, TENO schemes do  not directly use Eq.\eqref{eq:TENO1} to give the final weights. Instead,
the smoothness measurement in Eq.\eqref{eq:TENO1} is normalised at first, i.e.
 \begin{equation} \label{eq:norm}
\chi_k=\frac{\gamma_k}{\sum_{k=0}^{K-3}\gamma_k},
\end{equation}
\noindent and then a cut-off procedure is defined as
\begin{equation} \label{eq:seclect}
\delta_k=
\begin{cases}
\begin{matrix}
0, & \text{if} \quad \chi_k < C_T, \\
1, & \text{otherwise}.
\end{matrix}
\end{cases}
\end{equation}
Finally, the weights of TENO scheme for Eq. \eqref{eq:weighted} are defined by a normalizing procedure
\begin{equation} \label{eq:TENO_W}
\omega_k^{(T)}=\frac{d_k\delta_k}{\sum_{k=0}^{K-3}d_k\delta_k},
\end{equation}
\noindent where the optimal weights are utilised without rescaling, and only   the stencil  deemed to be containing discontinuity is   removed
from the final reconstruction completely.
Therefore, the numerical robustness of TENO scheme can be ensured, and the optimal weight, $d_k$, as well as the accuracy and spectral properties is fully recovered in smooth regions.

$C_T$ is also an effective and a direct mean to control the spectral properties of TENO scheme for a specific problem, e.g. compressible turbulence simulations. Moreover,
Haimovich and Frankel \cite{Haimovich2017} has conducted a series of numerical cases, in which the TENO solution with $C_T=10^{-3}$ is still superior in comparison to the WENO-Z solution. Fu et al. \cite{Fu2018} had also introduced an adaptive strategy of $C_T$,  minimizing numerical dissipation for high-wavenumber fluctuations.
In this article, since  the effect  of using this  parameter is not of the primary concern,  $C_T=10^{-5}$ is directly used for the fifth-order TENO scheme and the  method presented in this work, without specific investigation.


\section{Exploring TENO schemes as shock-detectors} \label{sec:new}

As introduced in \cite{Fu2018}, TENO schemes can be acting as a shock-detection-stencil-selection procedure.
In \cite{Fu2018}, six-point and eight-point TENO schemes are discussed,  and
the hierarchical voting strategy reduces the number of possible combinations in the stencil-selection process.
 In this section, the five-point TENO scheme is discussed following the idea in \cite{Fu2018}, to facilitate the introduction in the following sections.

\begin{figure}[h!t]
\begin{center}
\subfigure[\label{fig:f:stencils1}{A central stencil crossing discontinuity}]{
\resizebox*{5.5cm}{!}{\includegraphics{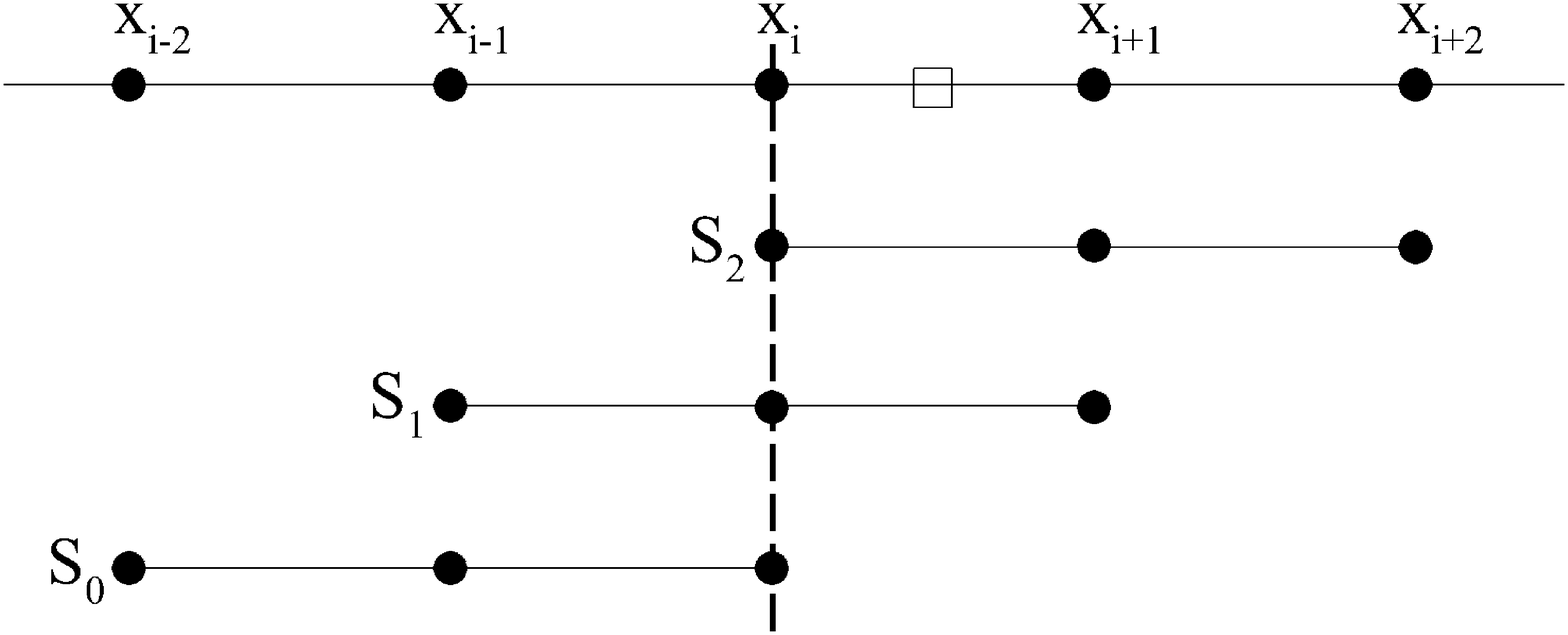}}}
\subfigure[\label{fig:f:stencils2}{Two candidate stencils crossing discontinuity}]{
\resizebox*{5.5cm}{!}{\includegraphics{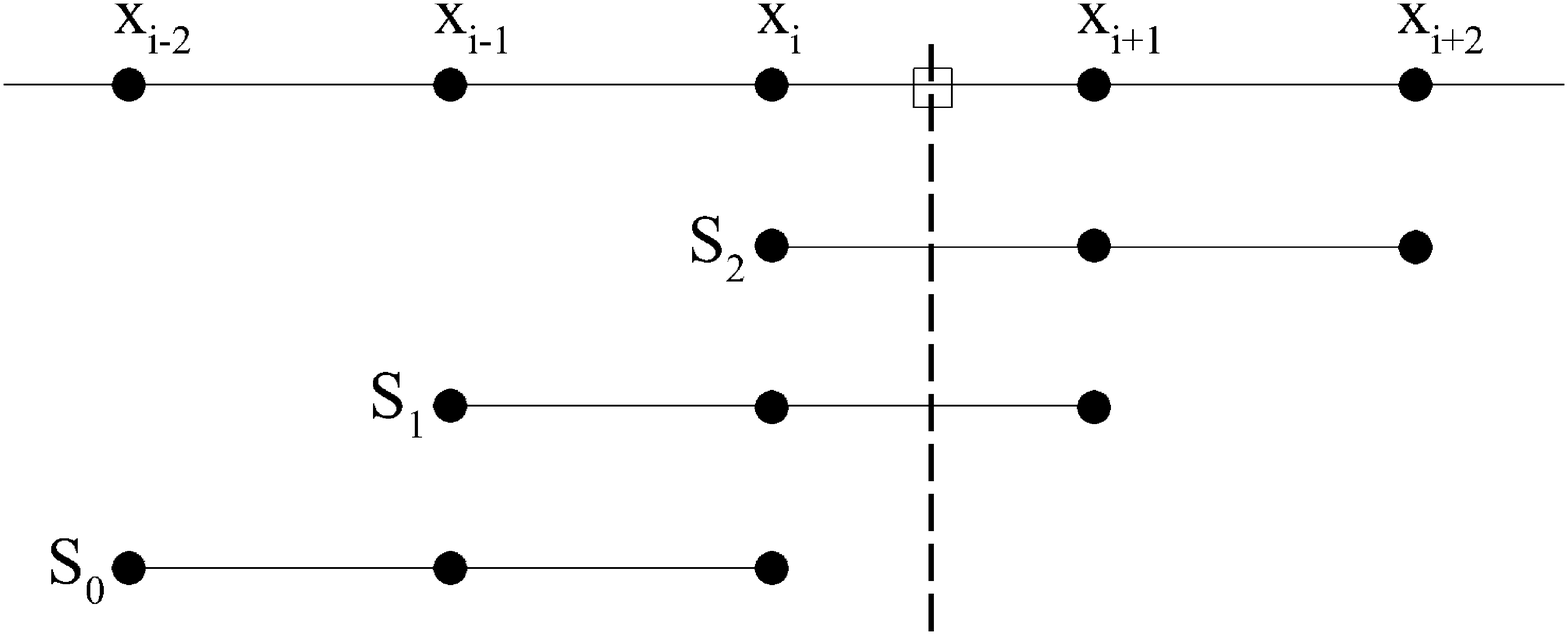}}}
\caption{\label{fig:f:stencils} Candidate stencils of a five-point (W/T)ENO scheme in the spatial approximation at $x_{i+\frac{1}{2}}$. }
\end{center}
\end{figure}

A WENO scheme uses continuous nonlinear weight to calculate convex combinations. Therefore, there are infinite possible combinations since the weights are calculated by continuous   function. Whereas,   TENO schemes in fact only implement several combination of the candidate stencils, by the nonlinear ENO-like  stencil-selection procedure in Eq.\eqref{eq:seclect}.
For the fifth-order TENO scheme, which has three stencils as shown in Fig.\ref{fig:f:stencils}, potential combinations include $\{\text{S}_0, \text{S}_1,\text{S}_2\}$, $\{\text{S}_0, \text{S}_1\}$, $\{\text{S}_1,\text{S}_2\}$, $\{\text{S}_0,\text{S}_2\}$, and $\{\text{S}_k\}$  if other two stencils are both crossing discontinuity, as shown  in Fig.\ref{fig:f:stencils2}. Especially, for each candidate stencil $\text{S}_k$, there are only two potential choices of the weight, i.e. $\{d_k, 0\}$, before the normalization procedure in Eq.\eqref{eq:TENO_W}.

Obviously, 
  the numerical flux as Eq.\eqref{eq:weighted} is  constructed based on a combination of candidate stencils
and    can be represented as a single high-order polynomial,  resulting the fifth-order upwind central scheme ($\{\text{S}_0, \text{S}_1,\text{S}_2\}$) or the four-point   schemes ($\{\text{S}_0,\text{S}_1\}$ or $\{\text{S}_1,\text{S}_2\}$), etc. Numerically, using these single high-order polynomial reconstructions is equivalent to the original TENO scheme, discarding the weighting procedure completely. 
Therefore, the equivalent numerical fluxes of the fifth-order TENO scheme evaluated at $x_{i+1/2}$  are given as
\begin{equation} \label{eq:unif1}
 \hat{f}_{m,i+1/2}^*=   a_{m,i-2}f_{i-2}+a_{m,i-1}f_{i-1} +a_{m,i}f_{i}+a_{m,i+1}f_{i+1}+a_{m,i+2}f_{i+2},
 \end{equation}
 where the subscript \textit{m} is used for numbering the polynomials and distinguishing to the stencils using subscript \textit{k} which relates to the local smoothness indicators. The coefficients of these polynomials
 are given in Table \ref{table:co1}.

\begin{table}
\scriptsize
\centering
\caption{ The coefficients of the equivalent single polynomial spatial reconstructions.}\label{table:co1}
\begin{tabular}{cccccccc}
\hline
\text{if} $\delta_{0,1,2}=$ &$\hat{f}_{m,i+1/2}^*$& S$_m^*$  & $a_{m,i-2}$  &  $a_{m,i-1}$ & $a_{m,i}$ &  $a_{m,i+1}$ &  $a_{m,i+2}$    \\ \hline
1,1,1                       &$\hat{f}_{0,i+1/2}^*$& S$_0^*$  & 1/30         &  -13/60    &    47/60   & 9/20 &  -1/20     \\
0,1,1                       &$\hat{f}_{1,i+1/2}^*$& S$_1^*$  & 0            & -1/9       &  2/3       & 1/2  &   -1/18     \\
1,1,0                       &$\hat{f}_{2,i+1/2}^*$& S$_2^*$  & 1/21         & -13/42     &   41/42    & 2/7  & 0                \\
0,0,1                       &$\hat{f}_{3,i+1/2}^*$& S$_3^*$  & 0            & 0          &  1/3       & 5/6  & -1/6      \\
0,1,0                       &$\hat{f}_{4,i+1/2}^*$& S$_4^*$  & 0            &  -1/6      & 5/6        & 1/3  &   0   \\
1,0,0                       &$\hat{f}_{5,i+1/2}^*$& S$_5^*$  & 1/3          & -7/6       &  11/6      & 0    & 0      \\
1,0,1                       &$\hat{f}_{6,i+1/2}^*$& S$_6^*$  &    1/12      & -7/24      &  17/24     & 5/8  & -1/8     \\
\hline
\end{tabular}
\end{table}

 \begin{figure}
 \centering
 \includegraphics[width=6cm]{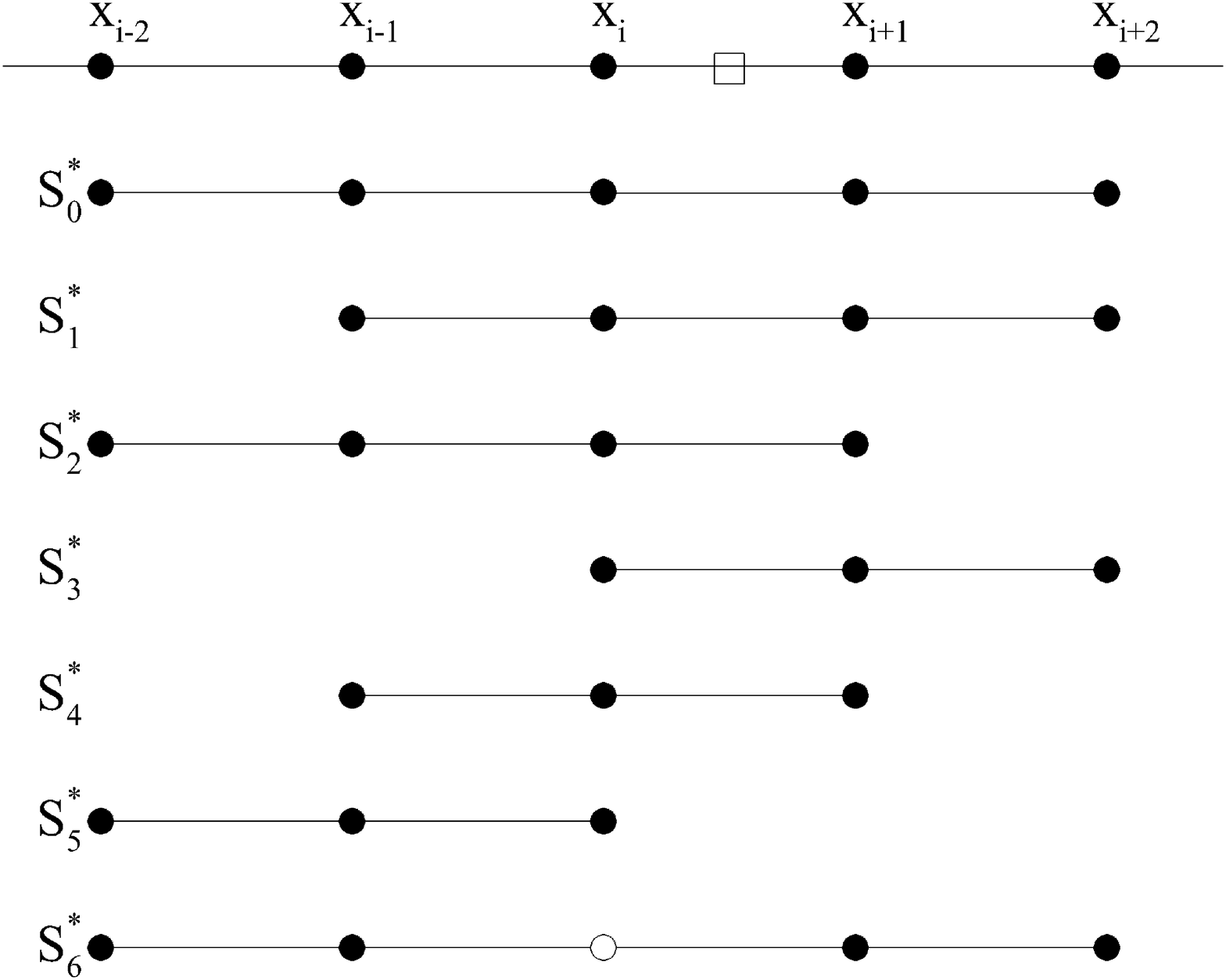}
 \caption{ \label{fig:f:re}
  Schematic of the equivalent candidate stencils of the fifth-order TENO scheme. The circle indicates that the stencil is not \textit{really} continuous.}
\end{figure}

Here, $\hat{f}_{m,i+1/2}^*$ is constructed on $S_m^*$ shown in Fig.\ref{fig:f:re}. Thereinto, $\hat{f}_{6,i+1/2}^*$ constructed on stencil $S_6^*$ is equivalent to the numerical flux constructed based on the combination of $\{\text{S}_0,\text{S}_2\}$, but it is not an actual continuous reconstruction or of fifth-order of accuracy.

 It should be mentioned explicitly, only one numerical flux $\hat{f}_{m,i+1/2}^*$ needs to be be selected in the spatial reconstruction, equivalently representing a weighted averaged numerical flux of the original TENO scheme.
 This explanation  brings flexibility to independently define the smoothness measurement and the spatial reconstructions, leading to more further possibilities. For example, the smoothness measurement can be improved or changed to improve discontinuity-detection capability, without changing the candidate high-order reconstructions. One may also apply a different scheme on a specific stencil, e.g. a central scheme on $S_1^*$, without changing the other polynomials. In general,
  TENO schemes can remove all the oscillating stencil(s)  and construct the  best polynomial (high-order of accuracy or high-resolution) that can be achieved on the known   smooth field.

\section{The presented method} \label{sec:method}

\subsection{The primary idea}

As well known and mentioned in the first section, achieving higher-order of spatial accuracy needs to increase DOFs in constructing polynomials. However, WENO schemes or TENO schemes cost significant more computation resources while increasing DOFs, as schematically shown in Fig.\ref{fig:f:7order}.
Therefore, it is straightforward to expect a simple but effective mean to achieve high numerical accuracy with low computation cost. In this subsection, we discuss the current methods and expect to find the potential of further improvement.

Preliminary, we can find that the smoothness indicators have two basic responsibilities, i.e. measuring local oscillation/smoothness and defining the contribution of each candidate stencil.  In WENO schemes, these two aspects are strongly coupled, since the specific   contribution of a candidate stencil  is calculated by using the continuous function of the local smoothness measurement.
More specifically, it is not enough to simply detect discontinuities, but the relative smoothness is also important to know which candidate stencil shall be more contributive in a WENO-family reconstruction.

Whereas, as aforementioned, TENO schemes  are  acting as a shock-detection-stencil-selection procedure. The responsibility of the smoothness measurement in TENO schemes is only to determine the location of the  captured discontinuity, and then the high-order optimal linear schemes can be applied wherever the stencils are smooth. More specifically, for TENO schemes, it is unnecessary to know which candidate stencil is more smooth or less smooth, as long as they are  indeed \textit{smooth}. It should be noticed again, the ENO-like stencil-selection procedure is essential to achieve this advantage.

Therefore, it is possible to reduce the computation cost if all the discontinuities can be determined in a more effective way, or wider  smooth field can be taken into account without using more complex smoothness measurement.


\subsection{Exploiting neighbouring smoothness information}
According to the former discussions, for TENO schemes, it is unnecessary to provide exact relative smoothness of each candidate stencil in constructing high-order polynomials.
Examining the discontinuity-detection issue in the whole spatial field, it could be found that all the discontinuities can be effectively detected by using the smoothness measurement of the fifth-order TENO scheme \cite{zhangcicp_2018}. In other word, the  smoothness measurement works sufficiently well to provide the smoothness information of the whole field, as least in those typical gas dynamic problems.
Therefore, if all the discontinuities are successfully detected, it might be possible to achieve arbitrary high-order interpolation as long as there is a sufficiently large smooth region.

\begin{figure}
 \centering
 \includegraphics[width=8cm]{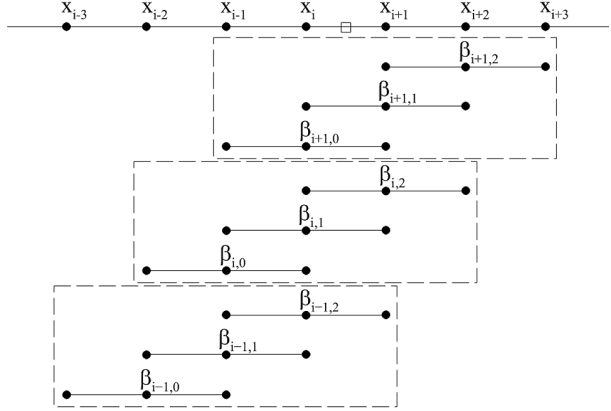}
 \caption{ \label{fig:f:stencil}
  The neighbouring smoothness measurement to be explored.}
\end{figure}

However, the smoothness information can not be directly used in current frameworks. As shown in Fig.\ref{fig:f:stencil}, one may notices that in a five-point WENO or TENO scheme approximating the variables at $x_{i+1/2}$, only the information  based on $\beta_{i,0}$, $\beta_{i,1}$ and $\beta_{i,2}$
can be applied to determine discontinuities or smooth field, and achieving  fifth-order  accuracy is  the best result in this context, \textit{even the neighbouring five-point full stencils are all smooth}.

It is obvious that the  three five-point full stencils in Fig.\ref{fig:f:stencil} cover the the same region as the seventh-order WENO or TENO scheme, and thus we can investigate this case.
Examining the three five-point full stencils in Fig.\ref{fig:f:stencil}, one may find that they are overlapping with each other. Therefore, it is highly possible that one discontinuity is detected by more than one five-point full stencil.
Examining the candidate stencils of the seventh-order TENO scheme, as shown in Fig.\ref{fig:f:7order2}, it is obvious that the candidate stencils of the fifth-order TENO scheme are included. As   known, the leading term of the local smoothness indicator $\beta_{i,k}$ is of $\mathcal{O}(\Delta x^2)$ if the corresponding stencil is smooth. Otherwise, if a discontinuity is located in a certain stencil, the corresponding $\beta_{i,k}$ is of $\mathcal{O}(1)$, invoking the ENO-like stencil-selection procedure.

Therefore, in order to show the capability of the neighbouring smoothness measurement in providing effective information for constructing a higher-order approximation at $x_{i+1/2}$, following remarks are given.

\noindent \textbf{Remark 1.} Examining Figs.\ref{fig:f:7order2} and \ref{fig:f:stencil}, if a discontinuity is located within the range of the central five-point full stencil,  one of the local smoothness indicators of the five-point full stencil or at least one of the local smoothness indicators of the seventh-point full stencil will  be of $\mathcal{O}(1)$. If  a discontinuity is located within the range of the seven-point full stencil, but outside the range of the central five-point full stencil, one of the local smoothness indicators of the neighbouring five-point full stencil or the seventh-point full stencil will  be of $\mathcal{O}(1)$. Eventually, the discontinuity can be located accurately by using either the three five-point smoothness measurements in Fig.\ref{fig:f:stencil} or the seven-point smoothness measurement in Fig.\ref{fig:f:7order2}.
$\Box$

\noindent \textbf{Remark 2.} Regarding to attaining designed order of accuracy, the ENO-like selection procedure helps to give a relaxed condition \cite{Fu2017} as
 \begin{equation} \label{eq:order}
\frac{\tau_{i,5}}{\beta_{i,k}+\epsilon}=O(\Delta x^s), \quad s>0,
\end{equation}
\noindent where $\tau_{i,5}=|\beta_{i,0}-\beta_{i,2}|$. This is the  condition required  for  TENO  schemes   to   restore the  formal  order  of   accuracy in   smooth   regions.   In fact, as aforementioned, using Eq.\eqref{eq:TENO1} significantly enhances  discontinuity-detection   capability of the smoothness measurement, and the ENO-like stencil-selection procedure directly applies those optimal linear schemes, except for those oscillating stencils where  discontinuities are located, i.e. where $\chi_k<C_t$.
$\Box$

\noindent \textbf{Remark 3.} In hyperbolic system equations, e.g. Euler equations, characteristic-wise reconstruction is necessary to avoid oscillation \cite{Balsara2000}. Therefore, the matrix $\mathbf{L}$ which is the left eigenvectors of the Jacobian matrix $\mathbf{A} =\frac{\partial \mathbf{F}}{\partial \mathbf{U}}$ should be used to calculate the characteristic variables, i.e.
 \begin{equation}
 \mathbf{Q}_l=\mathbf{L}_{i+1/2}\cdot \mathbf{U}_l, \quad i-2\le l \le i+2.
\end{equation}
\noindent To be clear, here $\mathbf{F}$ and $\mathbf{U}$ are the numerical flux and the conservative variables of the hyperbolic system equations, respectively.

It is unnecessary to detail the characteristic-wise reconstruction procedure. However, it does need to notice that the three five-point smoothness measurements in Fig.\ref{fig:f:stencil} are calculated based on the characteristic variables calculated using different Jacobian matrices, i.e. $\mathbf{A}_{i-1/2}$, $\mathbf{A}_{i+1/2}$ and $\mathbf{A}_{i+3/2}$, respectively. Luckily, in smooth field, the difference of linearised matrix $\mathbf{A}$, i.e. $\Delta\mathbf{A}$,  is of $\mathcal{O}(\Delta x)$, and thus the characteristic variables calculated using  $\mathbf{L}'=\mathbf{L}+\Delta \mathbf{L}$ does not change the essential property of the smoothness measurement.
$\Box$




%
%

Eventually, for the one-dimensional field as shown in Fig.\ref{fig:f:stencil},   two neighbouring stencils  are also utilised  in calculating the numerical flux at $x_{i+\frac{1}{2}}$.
A binary vector storing the information of local smoothness  is then  given as
 \begin{equation} \label{eq:vec}
\Delta_i=(\delta_{i-1,0},\delta_{i,0},\delta_{i,1},\delta_{i,2},\delta_{i+1,2}).
 \end{equation}
\noindent For example, if the candidate stencils are all smooth, the vector should be $\Delta=(1,1,1,1,1)$. Otherwise, one element which is equal to zero  indicates a discontinuity  has been detected. Of course, more than one discontinuity is possible to be detected. By using this information, following higher-order reconstructions  can be given.

\subsection{Higher-order spatial reconstructions}

In the last subsection, based on the smoothness measurement of the five-point TENO scheme, a method  providing extra smoothness information is given, and thus higher-order polynomials can be used in the smooth region.
In \cite{Fu2018}, predefined optimal linear
schemes are implemented as candidate spatial reconstructions. A similar idea is given as follows.

Here,   up to seventh-order candidate reconstructions  are given, directly using   two   neighbouring points of five-point  schemes.
Similar to Eq.\eqref{eq:unif1}, all the numerical fluxes evaluated at $x_{i+1/2}$  are given as an unified form,
\begin{equation}
\hat{f}_{m,i+1/2}^{**}= \sum_{l=i-3}^{i+3}a_{m,l}f_l,   
\end{equation}
and the coefficients are given in Table \ref{table:co2}.

\begin{table}
\scriptsize
\centering
\caption{ The coefficients of the numerical flux functions of the presented method.}\label{table:co2}
\begin{tabular}{cccccccc}
\hline
$\hat{f}_{m,i+1/2}^{**}$&$a_{m,i-3}$&$a_{m,i-2}$&$a_{m,i-1}$&$a_{m,i}$&$a_{m,i+1}$&$a_{m,i+2}$&$a_{m,i+3}$   \\ \hline
$\hat{f}_{0,i+1/2}^{**}$& -1/140    &   5/84    &  -101/420 &  319/420& 107/210   & -19/210   & 1/105  \\
$\hat{f}_{1,i+1/2}^{**}$&  0        &  1/60     & -2/15     &  37/60  & 37/60     & -2/15     &  1/60    \\
$\hat{f}_{2,i+1/2}^{**}$&   -1/60   & 7/60      & -23/60    &  19/20  & 11/30     & -1/30     &  0           \\
$\hat{f}_{3,i+1/2}^{**}$&  0        & 0         & -1/20     &  9/20   & 47/60     &   -13/60  & 1/30  \\
$\hat{f}_{4,i+1/2}^{**}$& 0         & 1/30      & -13/60    &  47/60  &  9/20     & -1/20     &0    \\
$\hat{f}_{5,i+1/2}^{**}$& -1/20     & 17/60     & -43/60    &  77/60  & 1/5       & 0         &0   \\
$\hat{f}_{6,i+1/2}^{**}$&0          &0          & 0         &  1/4    & 13/12     & -5/12     &1/12    \\
$\hat{f}_{7,i+1/2}^{**}$&0          &0          & -1/12     &  7/12   & 7/12      & -1/12     &0    \\
$\hat{f}_{8,i+1/2}^{**}$&0          &1/12       & -5/12     &  13/12  & 1/4       & 0         &0    \\
$\hat{f}_{9,i+1/2}^{**}$&-1/4       &13/12      & -23/12    &  25/12  & 0         & 0         &0    \\
$\hat{f}_{10,i+1/2}^{**}$&0          &0          &0    & 1/3       & 5/6  & -1/6     &0    \\
$\hat{f}_{11,i+1/2}^{**}$&0          &0       &  -1/6      & 5/6        & 1/3        & 0         &0    \\
$\hat{f}_{12,i+1/2}^{**}$&0       &1/3          & -7/6       &  11/6  & 0         & 0         &0    \\
\hline
\end{tabular}
\end{table}

Examining these coefficients,  compared with the stencils in Fig.\ref{fig:f:re}, we can simply find that apart from the original high-order representatives, the presented method exploits several   higher-order polynomials, including the seventh-order polynomial $\hat{f}_{0,i+1/2}^{**}$, sixth-order polynomials $\hat{f}_{1,i+1/2}^{**}$ and $\hat{f}_{2,i+1/2}^{**}$, and also two more fifth-order polynomials, i.e. $\hat{f}_{3,i+1/2}^{**}$ and $\hat{f}_{5,i+1/2}^{**}$, as well as two forth-order polynomials $\hat{f}_{6,i+1/2}^{**}$ and $\hat{f}_{9,i+1/2}^{**}$. All these newly introduced high-order reconstructions are constructed with using the two extra neighbouring points. At the meantime, the coefficients of $\hat{f}_{7,i+1/2}^{**}$ and $\hat{f}_{8,i+1/2}^{**}$ are different to those forth-order polynomials in Table \ref{table:co1}, even they are using the same stencils. This is because the coefficients in  Table \ref{table:co1} are calculated using the optimal linear weights in Eq.\eqref{eq:opt},
after the ENO-like selection. Moreover, the stencil, S$_6^*$, is discarded since it is in fact crossing a central discontinuity.

Eventually, we can find the basic features of the presented method. Firstly, all the final reconstructions do not cross any discontinuity. In fact, WENO schemes or the TENO schemes before \cite{Fu2018} are possible to calculate the final approximation using the numerical fluxes in both sides of a discontinuity. For example, in Fig.\ref{fig:f:stencils1}, the central stencil could be discarded but the other two stencils could be applied. Secondly, the presented method is also suitable to deal with multiple closely located shocklets, because the polynomial constructed on a smaller stencil could be applied in such case.

\subsection{The stencil-selection procedure}

The TENO framework allows the separation of high-order spatial reconstructions and smoothness measurement.
For calculating the high-order polynomials introduced in the last subsection, the smoothness information should be provided at first, which means that the local smoothness indicators, $\beta_{i,k}$, as well as the global smoothness indicators $\tau_{i,5}$, should be readily available before the spatial reconstructions, providing the binary vector in Eq.\eqref{eq:vec}. In this work, the method calculating the local and global smoothness indicators is the same as that of the fifth-order TENO scheme, and thus the binary vector storing smoothness information can be calculated trivially.
 The relation between this binary vector and the implementation of the high-order reconstructions in Table \ref{table:co2} is then given in Table \ref{table:co3}.

\begin{table}
\scriptsize
\centering
\caption{ The coefficients of the numerical flux functions of the presented method. (* indicates 1 or 0)}\label{table:co3}
\begin{tabular}{cccc}
\hline
$\Delta_i=$&$\hat{f}_{m,i+1/2}^{**}$ & $\Delta_i=$&$\hat{f}_{m,i+1/2}^{**}$   \\ \hline
$(1,1,1,1,1)$    &$\hat{f}_{0,i+1/2}^{**}$   &   $(*,0,1,1,0)$  &$\hat{f}_{7,i+1/2}^{**}$         \\
$(0,1,1,1,1)$  &$\hat{f}_{1,i+1/2}^{**}$  &  $(0,1,1,0,*)$  &$\hat{f}_{8,i+1/2}^{**}$          \\
$(1,1,1,1,0)$  &$\hat{f}_{2,i+1/2}^{**}$ &     $(1,1,0,0,*)$  &$\hat{f}_{9,i+1/2}^{**}$      \\
$(*,0,1,1,1)$  &$\hat{f}_{3,i+1/2}^{**}$ &  $(*,*,0,1,0)$  &$\hat{f}_{10,i+1/2}^{**}$  \\
$(0,1,1,1,0)$  &$\hat{f}_{4,i+1/2}^{**}$ & $(*,0,1,0,*)$  &$\hat{f}_{11,i+1/2}^{**}$   \\
$(1,1,1,0,*)$  &$\hat{f}_{5,i+1/2}^{**}$ &  $(0,1,0,0,*)$  &$\hat{f}_{12,i+1/2}^{**}$  \\
$(*,*,0,1,1)$  &$\hat{f}_{6,i+1/2}^{**}$ &   $(*,0,0,0,*)$  &  no exist      \\
    \hline
\end{tabular}
\end{table}

It should be noticed that there are 2$^5$=32 possible combinations of Eq.\eqref{eq:vec}, and each of them is correspondent to a numerical flux reconstruction, except that the three central smoothness indicators can not be all oscillatory and reconstructions crossing discontinuity are completely avoided. Moreover, two situations need to be further explained. First, if $\Delta_i=(1,1,0,1,1)$, this is, theoretically,  available for using  $\hat{f}_{6,i+1/2}^{**}$ and $\hat{f}_{9,i+1/2}^{**}$. However, considering that the numerical flux at $x_{i+1/2}$ is to be evaluated,   $\hat{f}_{6,i+1/2}^{**}$ is used to make this evaluation an interpolation, instead of an extrapolation while using $\hat{f}_{9,i+1/2}^{**}$. Second, similar to the first case, $\Delta_i=(1,1,0,1,0)$  is correspondent to $\hat{f}_{10,i+1/2}^{**}$, instead of $\hat{f}_{9,i+1/2}^{**}$.

\begin{figure}
 \centering
 \includegraphics[width=10cm]{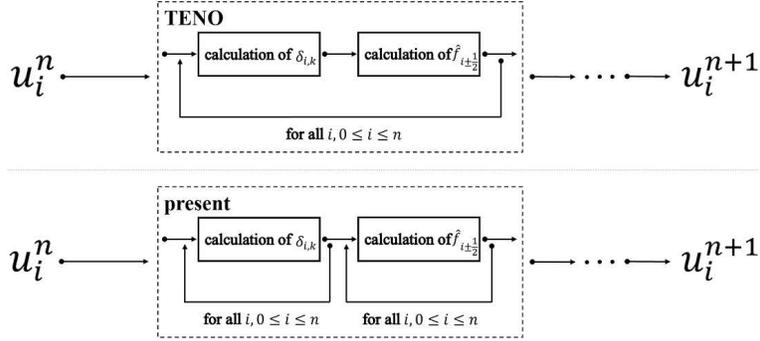}
 \caption{ \label{fig:f:compar}
  A minor difference between the procedures of a typical TENO scheme and the presented method.}
\end{figure}

By exploring the ENO-like selection procedure, a wider stencil can be used without amending  the original five-point smoothness measurement.
As mentioned in the beginning of this subsection,  the smoothness measurement of the whole computation domain should be calculated before the spatial reconstruction, for providing the \textit{neighbouring} smoothness information. Therefore, a minor difference is introduced in the procedure of the presented method, as briefly shown in Fig.\ref{fig:f:compar}, and the smoothness information should be stored in the calculation of $\hat{f}_{i\pm 1/2}$. The space complexity of the required storage is of $\mathcal{O}(n)$, and thus it will not be a barrier of applying the presented method.

%
%
%
\subsection{The parameter $\epsilon$}

The small parameter $\epsilon$ has shown great impact on the accuracy of WENO schemes \cite{Henrick2005,Don2013}, and it is also found that a small $\epsilon$ in the local smoothness indicator is possible to be overwhelmed by round-off error, leading to invoking the polynomial-selection procedure in unperturbed region, as in \cite{zhangcicp_2018}.

In this work, we do not specifically investigate this parameter since it does not affect the basic idea of the presented method. Therefore, considering that this parameter shall not suppress the effects of $\beta_k$ which is of $\mathcal{O}(\Delta x^2)$ in smooth field, we simply give a higher order infinitesimal
\begin{equation}
\epsilon=\Delta x^3.
\end{equation}

In \cite{Don2013}, the magnitude of $\epsilon$ was shown to be crucial for achieving designed order of accuracy at critical points. In this work, we do not further discuss this issue but only test the given formula, and the essential idea of this work can still be examined.
\section{Numerical results} \label{sec:results}

So far, the presented method is developed based on the smoothness measurement of the fifth-order TENO scheme, which was proved to be more accurate than WENO schemes. Therefore, we use the fifth-order TENO scheme  to produce baseline nonlinear approximations  in the following case, investigating both the accuracy and efficiency. The seventh-order WENO-Z scheme with $q=2$ is also applied in  discontinuity-capturing  simulations, for comparison.

\subsection{Approximate dispersion relation analysis}

Approximate dispersion relation analysis, or in short, ADR analysis, was developed by Pirozzoli \cite{Pirozzoli2006} and is suitable for analysing nonlinear numerical schemes. In this work, following the detailed instruction of Mao et al. \cite{Mao2015}, ADR analysis is performed at first.
Here, the fifth-order TENO scheme, the presented method and the corresponding fifth- and seventh-order linear schemes are used for comparison.

\begin{figure}[h!t]
\begin{center}
\subfigure[\label{fig:f:ADR_1}{dispersion}]{
\resizebox*{5.5cm}{!}{\includegraphics{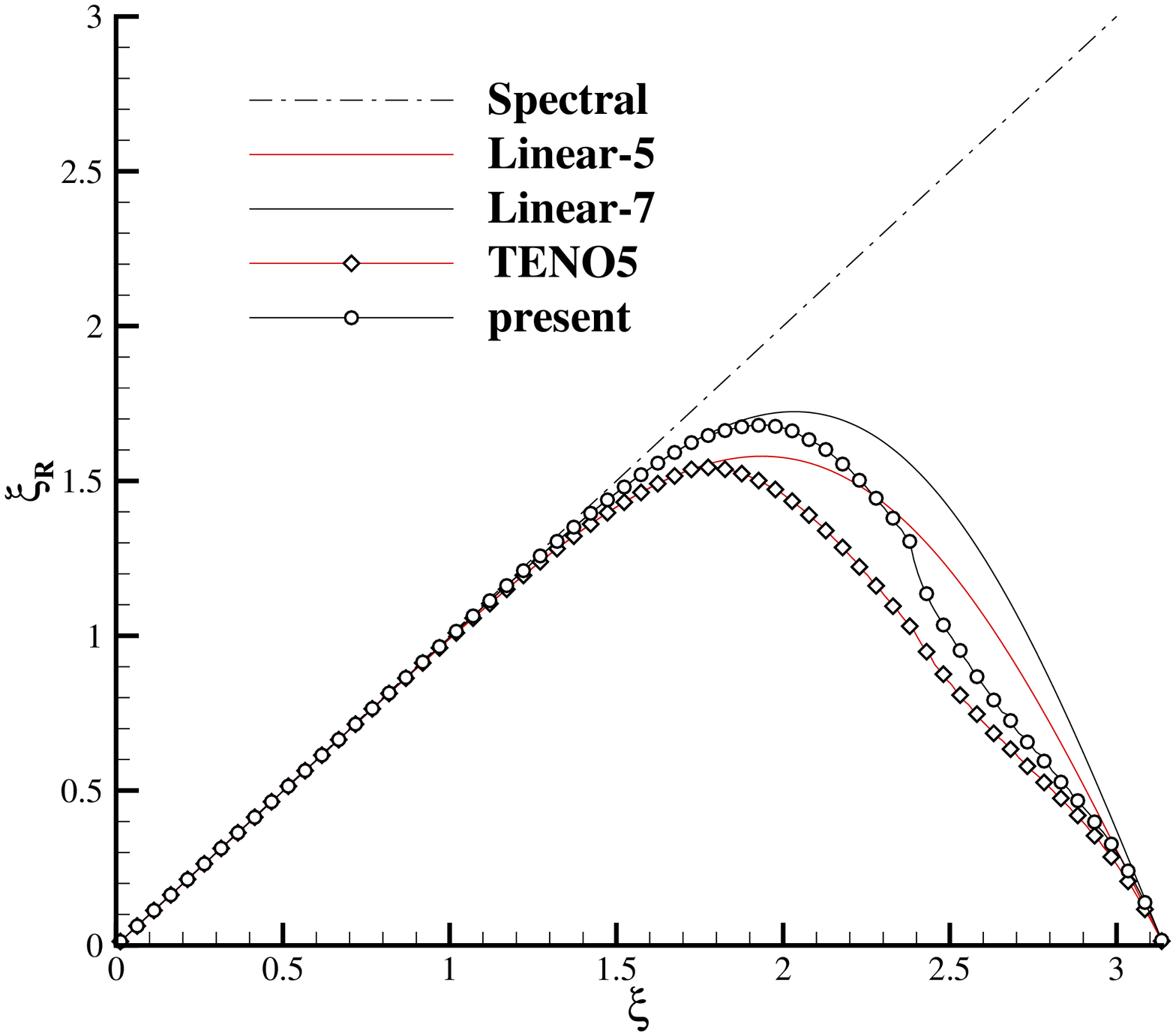}}}
\subfigure[\label{fig:f:ADR_2}{dispersion (zoom-in)}]{
\resizebox*{5.5cm}{!}{\includegraphics{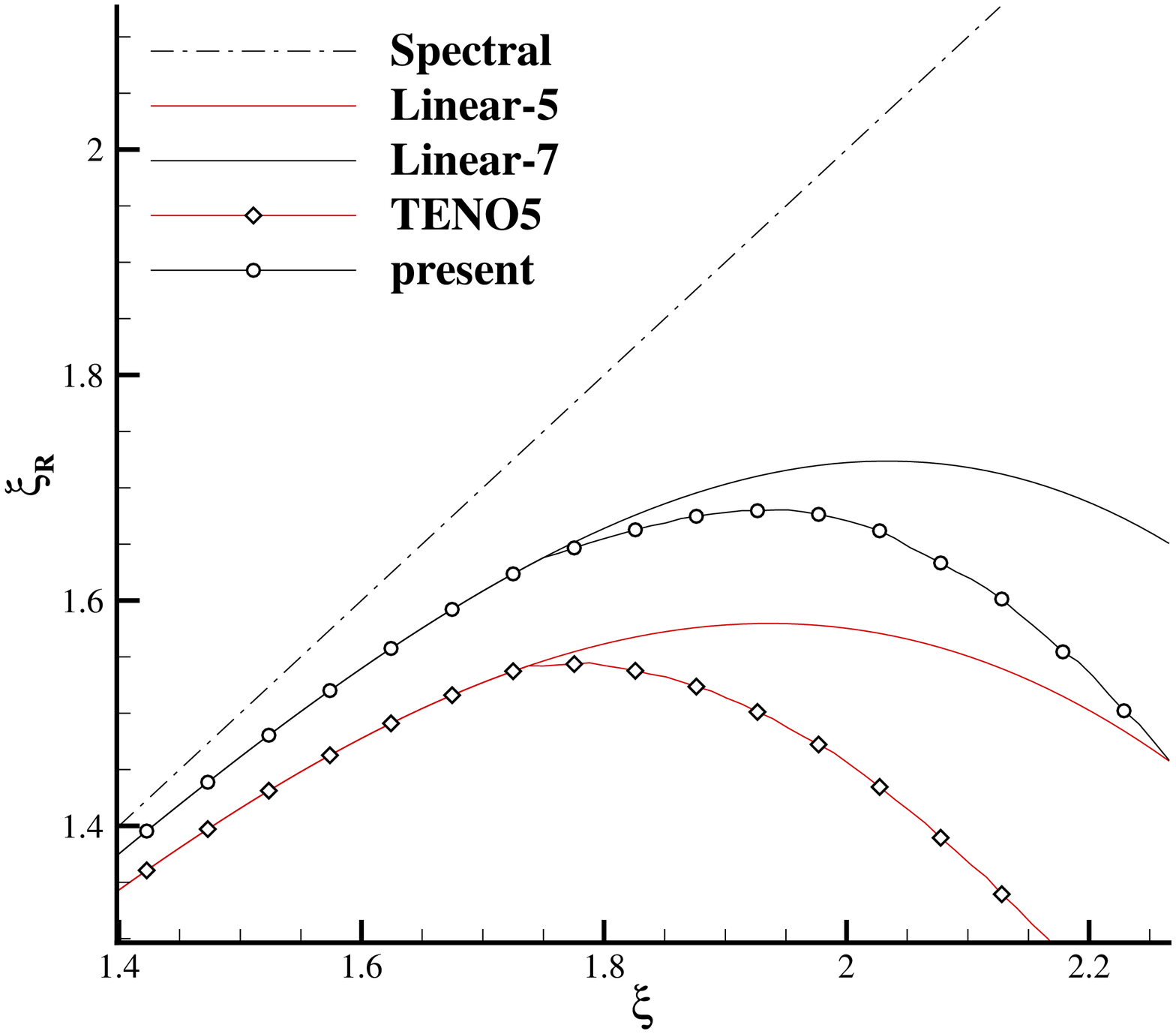}}}
\subfigure[\label{fig:f:ADR_3}{dissipation}]{
\resizebox*{5.5cm}{!}{\includegraphics{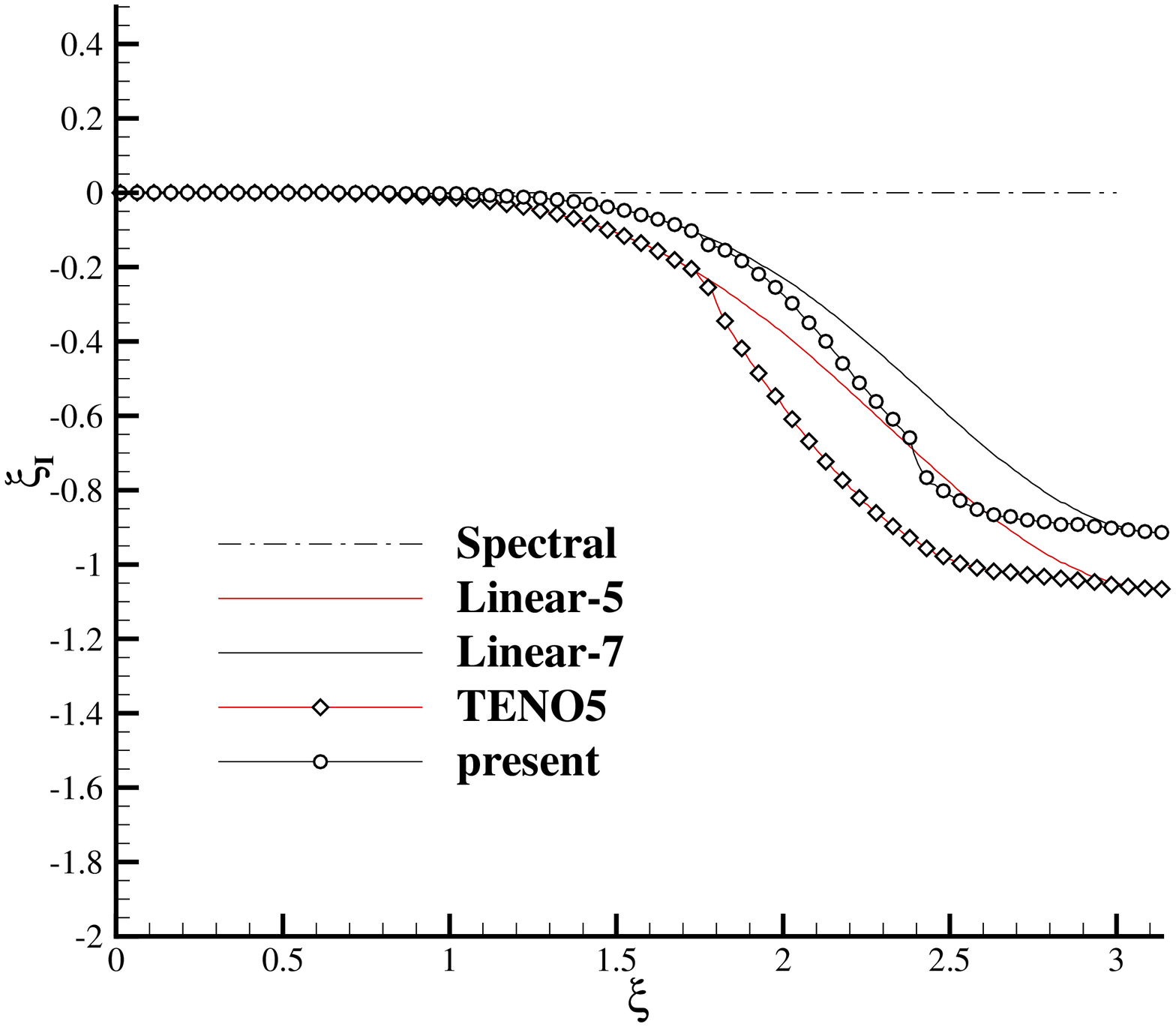}}}
\subfigure[\label{fig:f:ADR_4}{dissipation (zoom-in)}]{
\resizebox*{5.5cm}{!}{\includegraphics{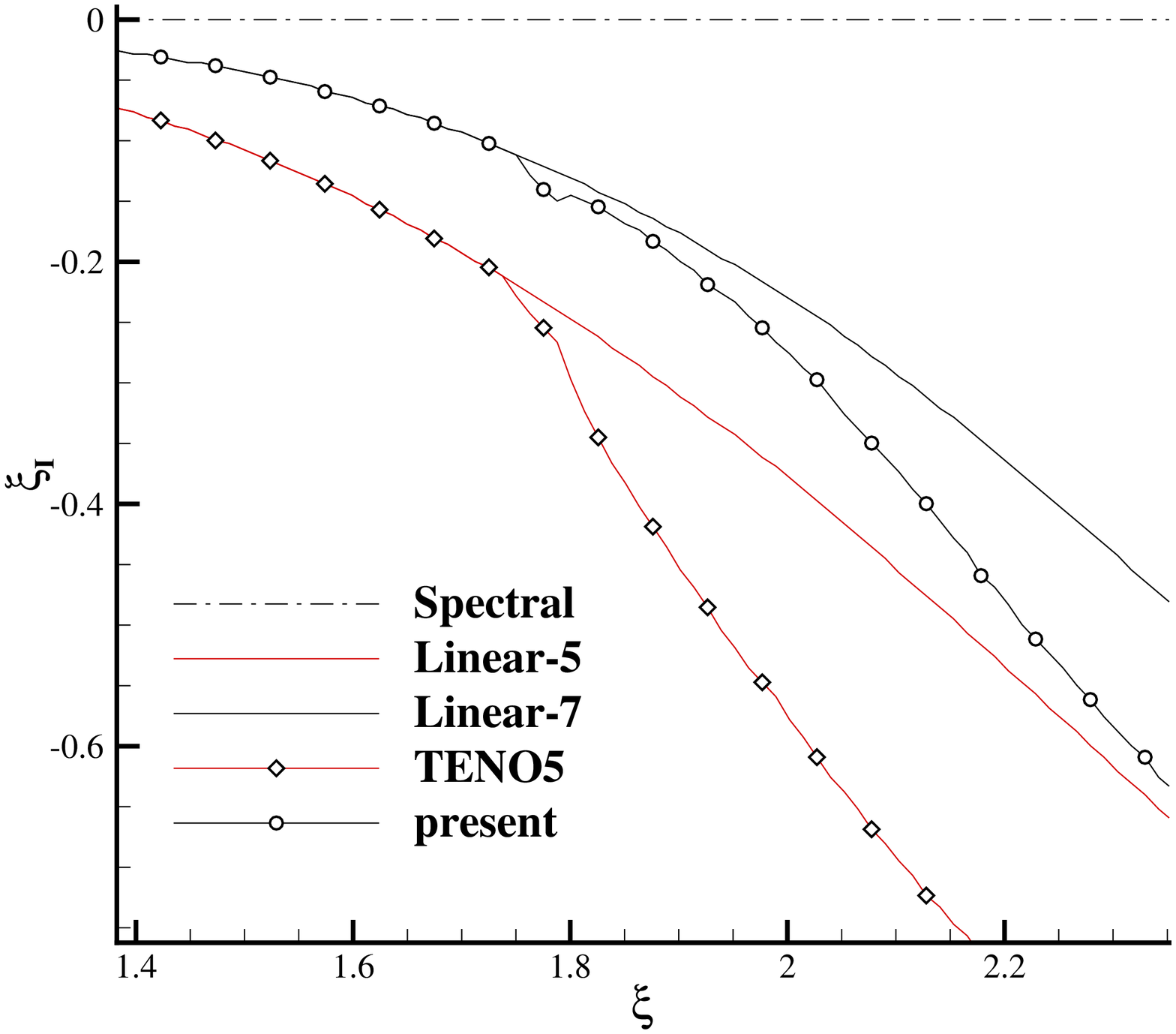}}}
\caption{\label{fig:f:ADR}The approximate dispersion and the dissipation properties of the presented method. }
\end{center}
\end{figure}

As shown in Fig.\ref{fig:f:ADR}, TENO schemes and the presented method show agreement with the background linear schemes in low and intermediate wave number. In high wave number region,  the results of the nonlinear schemes  deviate from  the results of the background linear schemes.
Especially, the presented method which uses the smoothness measurement of the fifth-order TENO scheme, deviates from the seventh-order linear scheme at the same wave number which causes deviation between the fifth-order TENO scheme and its background linear scheme. This result indicates that the smoothness measurement determines the resolvability of high frequency waves.
At the meantime, we can also find that the presented method shows significantly better numerical dissipation and dispersion at higher wave number region.  Therefore, exploiting the neighbouring information is effective to reduce the dissipation and dispersion error.

\subsection{One-dimensional linear advection problem}

TENO schemes recover the linear weights for simulating smooth waves, and thus while simulating a smooth field, TENO schemes are expected to behave like the linear schemes in terms of accuracy and resolution, which are the background five-point or seven-point linear scheme in the presented discussion.
The presented methods also need to recover the linear schemes. Especially, we need to prove that the simplifications do not deteriorate the performance of the schemes.
Here, the scalar linear advection problem having a smooth field or discontinuity is applied. The equation to be solved is
\begin{equation} \label{eq:LinAdv}
 \frac{\partial u}{\partial t}+ \frac{\partial u}{\partial x} =0,
\end{equation}
\noindent and periodic boundary condition is used to model the infinite one-dimensional scalar field.  The field is discretized using equidistributed  meshes, and the time steps are sufficiently refined to achieve the convergence of temporal solutions.

Three initial conditions are given as
\begin{equation} \label{eq:init_LinAdv}
u_1(x,0)=e^{-300(x-x_c)^2}, \quad x_c=0.5, \quad x\in[0,1),
\end{equation}
\begin{equation} \label{eq:init_LinAdv2}
u_2(x,0)=\sin(\pi x)^3, \quad x\in[0,2),
\end{equation}
\noindent and
\begin{equation} \label{eq:init_LinAdv3}
u_3(x,0)=
\begin{cases}
\begin{matrix}
& \frac{1}{6}\left(G(x,z-\delta)+G(x,z+\delta)+4G(x,z)\right),&\quad 0.2 \le x \le 0.4,\\
& 1, &\quad  0.6 \le x \le 0.8,\\
& 1-|10 (x-0.1)|,& \quad  1.0 \le x \le 1.2,\\
&\frac{1}{6}\left(F(x,a-\delta)+F(x,a+\delta)+4F(x,a)\right), &\quad 1.4 \le x \le 1.6, \\
 &0, &\quad \text{otherwise},
\end{matrix}
\end{cases}
\end{equation}
\noindent where the full domain is $x\in[-1,1)$. The two functions are defined as
\begin{equation} \label{eq:FG}
G(x,z)=\exp\left(-\beta(x-z)^2\right), \quad F(x,a)=\sqrt{\max(1-\alpha^2(x-a)^2,0)},
\end{equation}
\noindent where the coefficients are given as
\begin{equation} \label{eq:coof}
a=0.5,\quad z=0.7, \quad\delta=0.005, \quad\alpha=10,\quad\beta=\log(2)/(36\delta^2).
\end{equation}

\begin{figure}[h!t]
\begin{center}
\subfigure[\label{fig:f:LinAdv_1}{$L_1$ error}]{
\resizebox*{5.5cm}{!}{\includegraphics{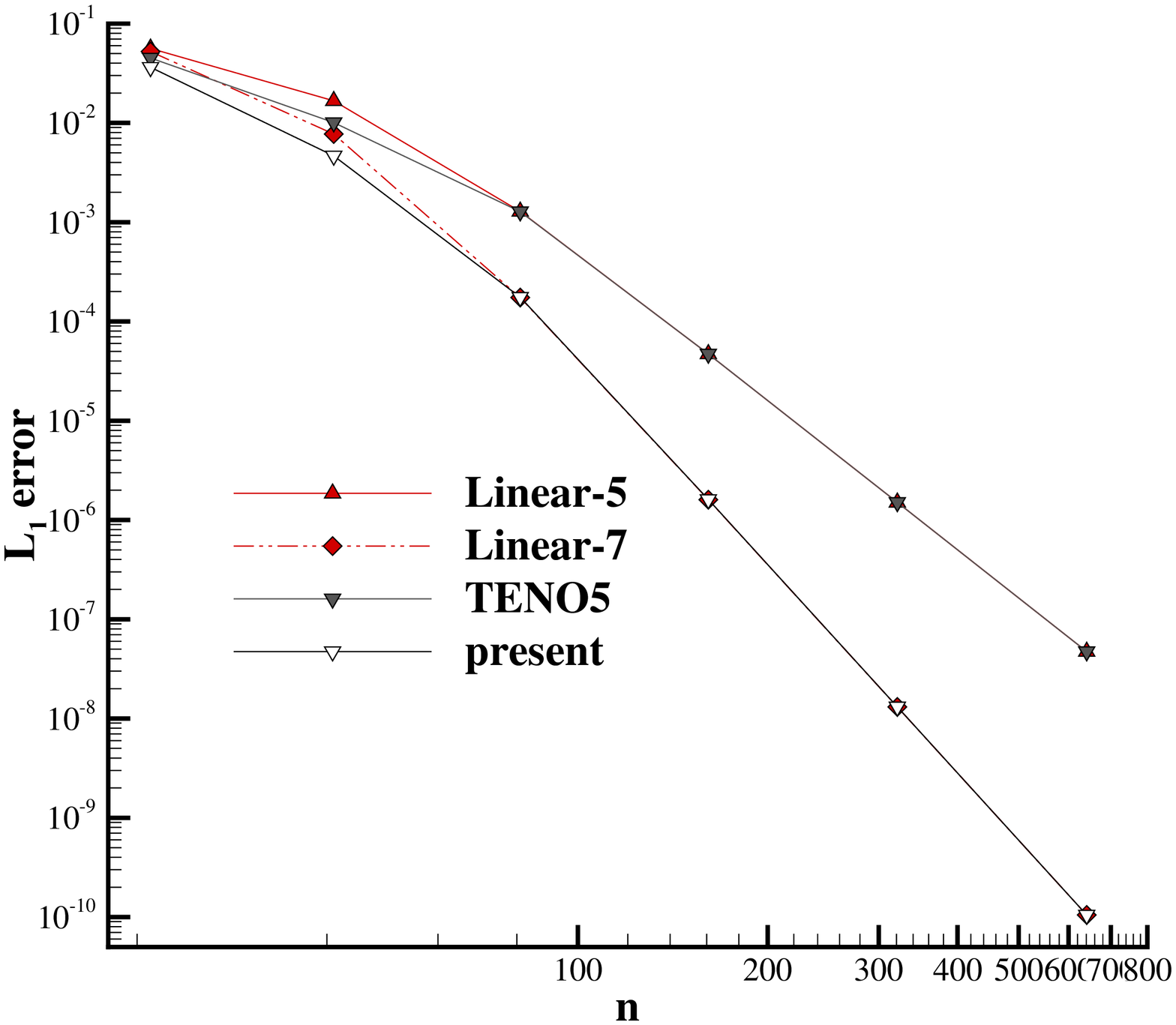}}}
\subfigure[\label{fig:f:LinAdv_2}{$L_{\infty}$ error}]{
\resizebox*{5.5cm}{!}{\includegraphics{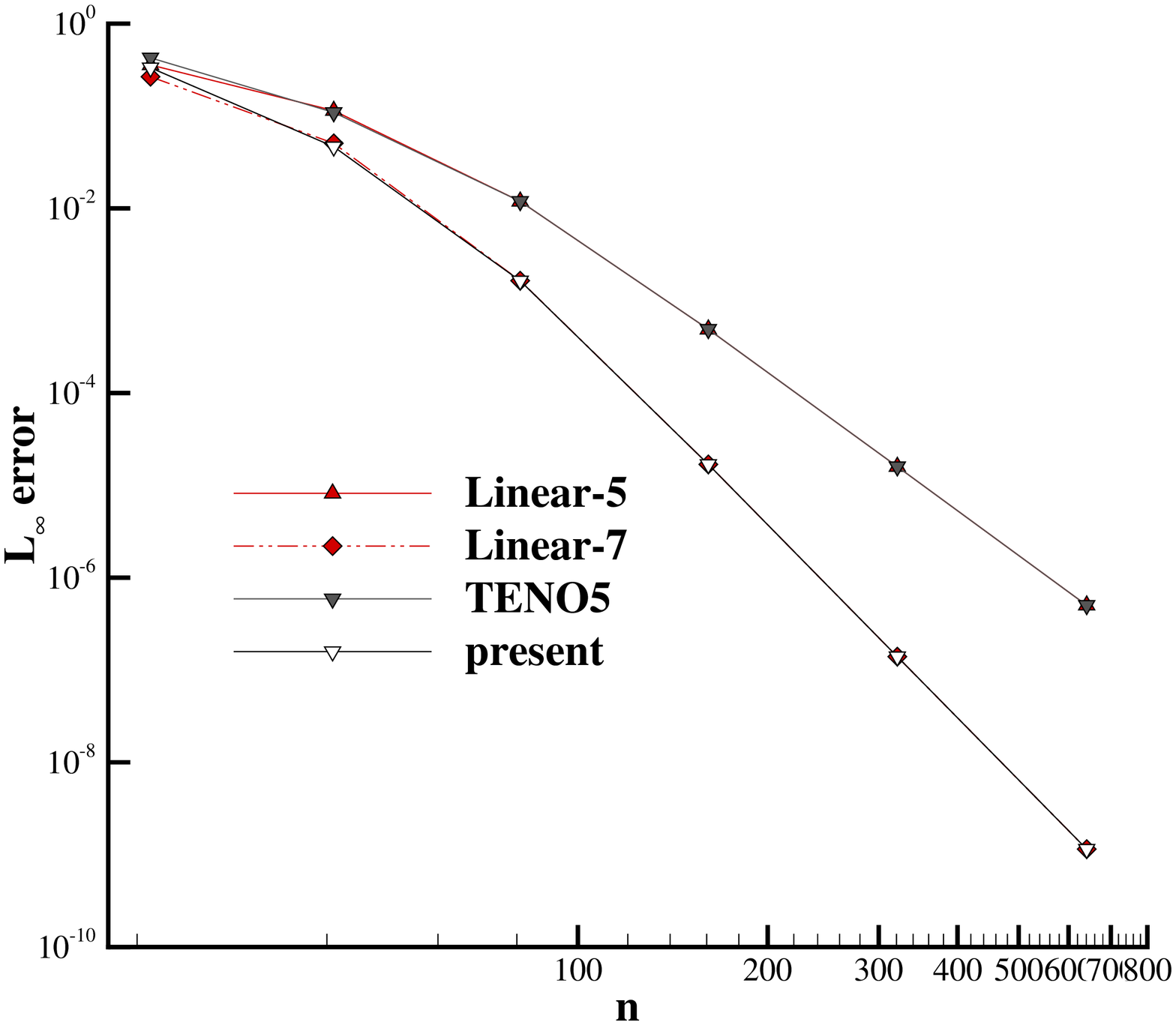}}}
\caption{\label{fig:f:LinAdv} The numerical error of the one-dimensional scalar advection solution: $u_1(x,0)$. }
\end{center}
\end{figure}

For $u_1(x,0)$, the numerical error is calculated at $t=1$,
which is one period of the solution. The results are shown in Fig.\ref{fig:f:LinAdv}, including the $L_1$ error and $L_{\infty}$ error. It can be found that
except on coarse discretizations, the resolution and accuracy of the TENO scheme and the presented method are the same as those of the background linear schemes. It should be noticed again the results of the background linear schemes are the best results which the nonlinear schemes can achieve.

\begin{figure}
 \centering
 \includegraphics[width=5.5cm]{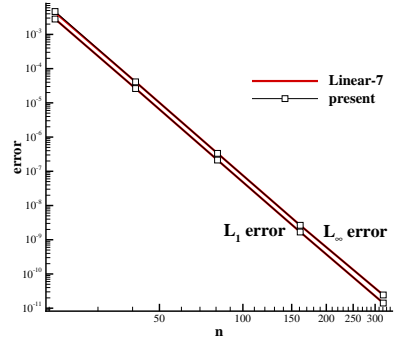}
 \caption{ \label{fig:f:linadv_2nd}
  The numerical error of the one-dimensional scalar advection solution: $u_2(x,0)$.}
\end{figure}

For $u_2(x,0)$, there are second-order critical points, where $\frac{\partial u}{\partial x}=\frac{\partial^2 u}{\partial x^2}=0$ and $\frac{\partial^3 u}{\partial x^3}\ne 0$. The error of the result at $t=1$ is shown in Fig.\ref{fig:f:linadv_2nd}. It can be found that the seventh-order background linear scheme  is again recovered by the presented method.  In general, by using these two smooth initial conditions in solving the linear advection equation, the accuracy of the presented method in smooth field including critical points is examined.

\begin{figure}[h!t]
\begin{center}
\subfigure[\label{fig:f:fourwave1}{}]{
\resizebox*{5.5cm}{!}{\includegraphics{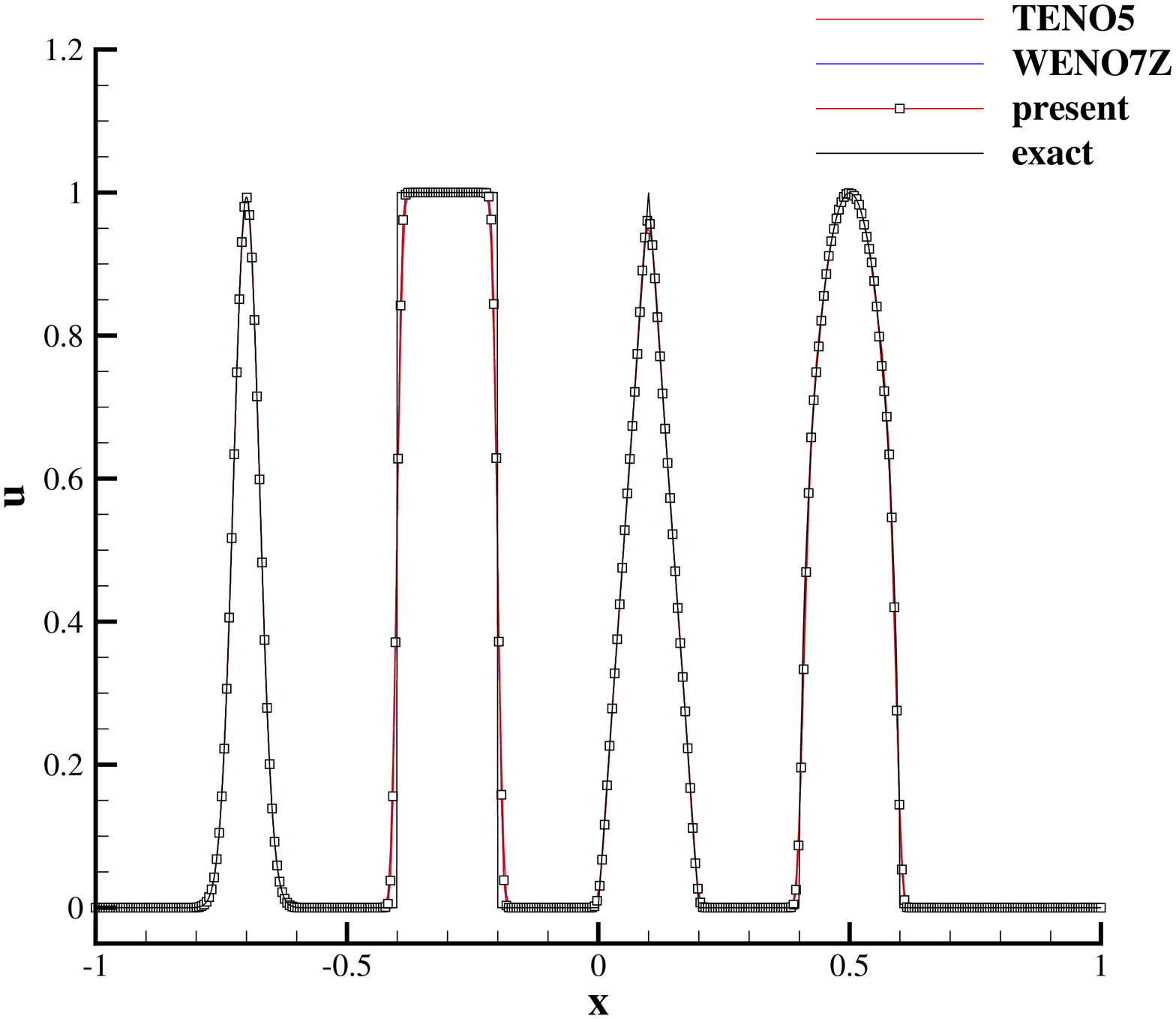}}}
\subfigure[\label{fig:f:fourwave2}{}]{
\resizebox*{5.5cm}{!}{\includegraphics{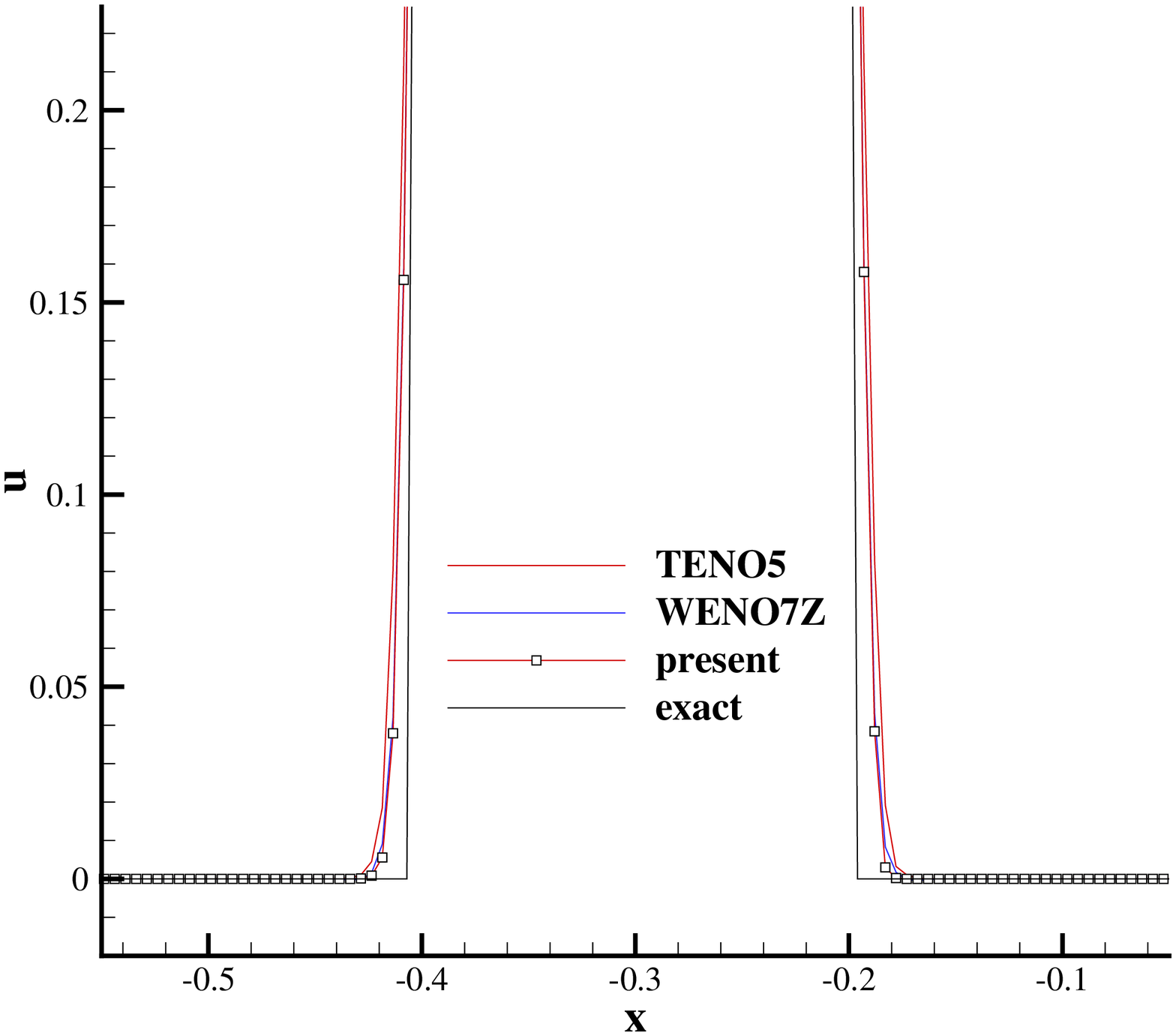}}}
\caption{\label{fig:f:fourwave} The numerical results of the one-dimensional   advection solution: $u_3(x,0)$. }
\end{center}
\end{figure}

By using the initial field $u_3(x,0)$, which  consists of several shapes containing
the combination of Gaussian, a square wave, a sharp triangle
wave and an half ellipse, is impossible to be solved by using the linear scheme. Especially, the presented method needs to be tested to find out whether it is oscillation-free near discontinuity. The equation is solved by using the fifth-order TENO scheme, the seventh-order WENO-Z scheme and the presented method on 400 grid points, and the results at $t=2$ are shown in Fig.\ref{fig:f:fourwave}.

Obviously, all the used methods show oscillation-free results. At the meantime, the presented method shows even slightly better resolution compared with the seventh-order  WENO-Z scheme.
\subsection{One-dimensional  inviscid Burgers equation}
The one-dimensional inviscid Burgers equation  is used to assess the actual order of accuracy of the proposed scheme when it is applied to a non-linear scalar equation. The governing equation is
\begin{equation}
\frac{\partial u}{\partial t} + \frac{\partial }{\partial x}\left(\frac{1}{2} u^2\right) = 0, \quad x\in [0,2),
\end{equation}
with periodic boundary-condition and the initial condition given by
\begin{equation}
u(x,0) = \frac{1}{2} + \text{sin}(\pi x).
\end{equation}

\begin{figure}[h!t]
\begin{center}
\subfigure[\label{fig:f:burgers0}{The solution on 160 grid points}]{
\resizebox*{5.5cm}{!}{\includegraphics{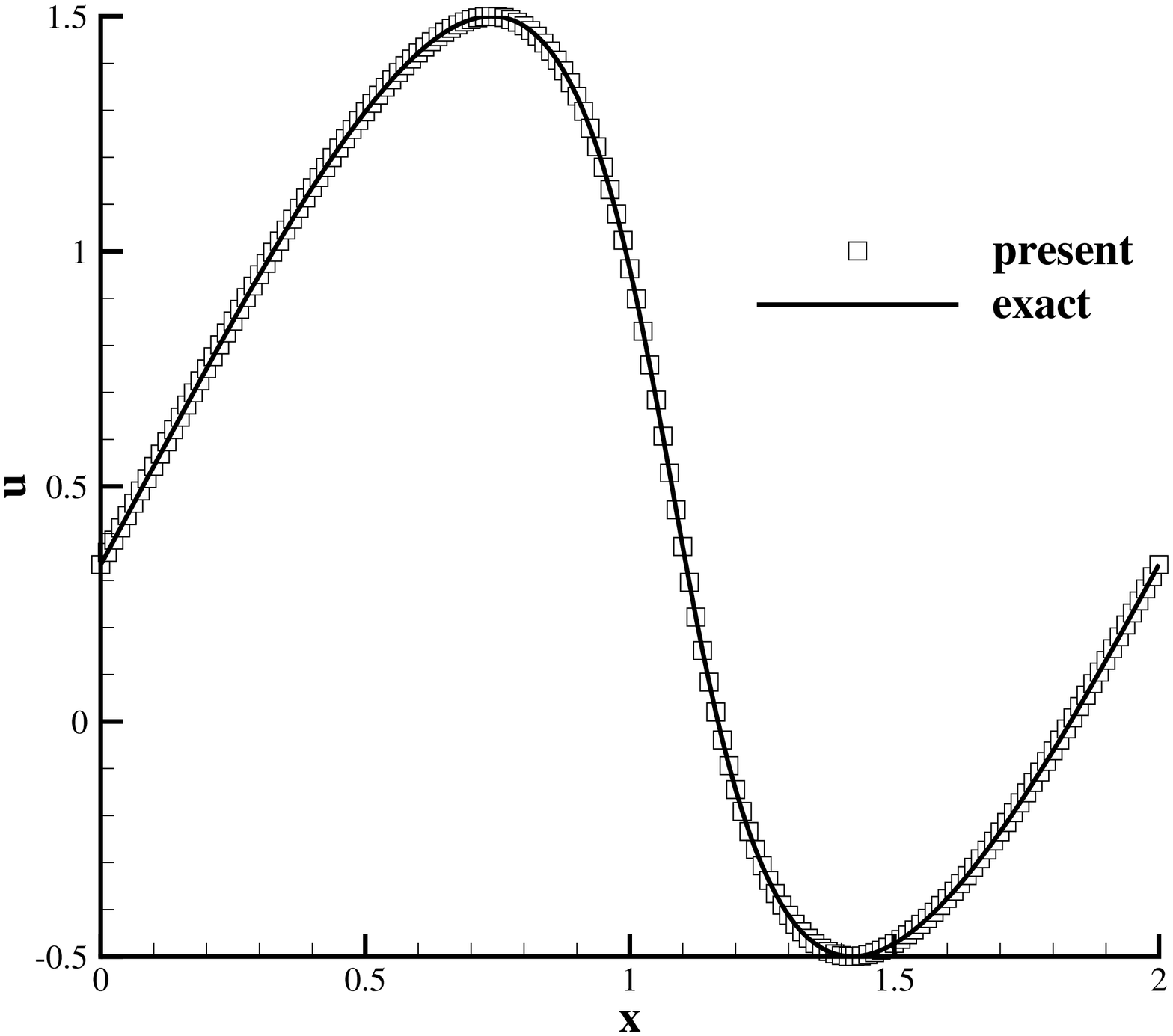}}}
\subfigure[\label{fig:f:burgers1}{Numerical error}]{
\resizebox*{5.5cm}{!}{\includegraphics{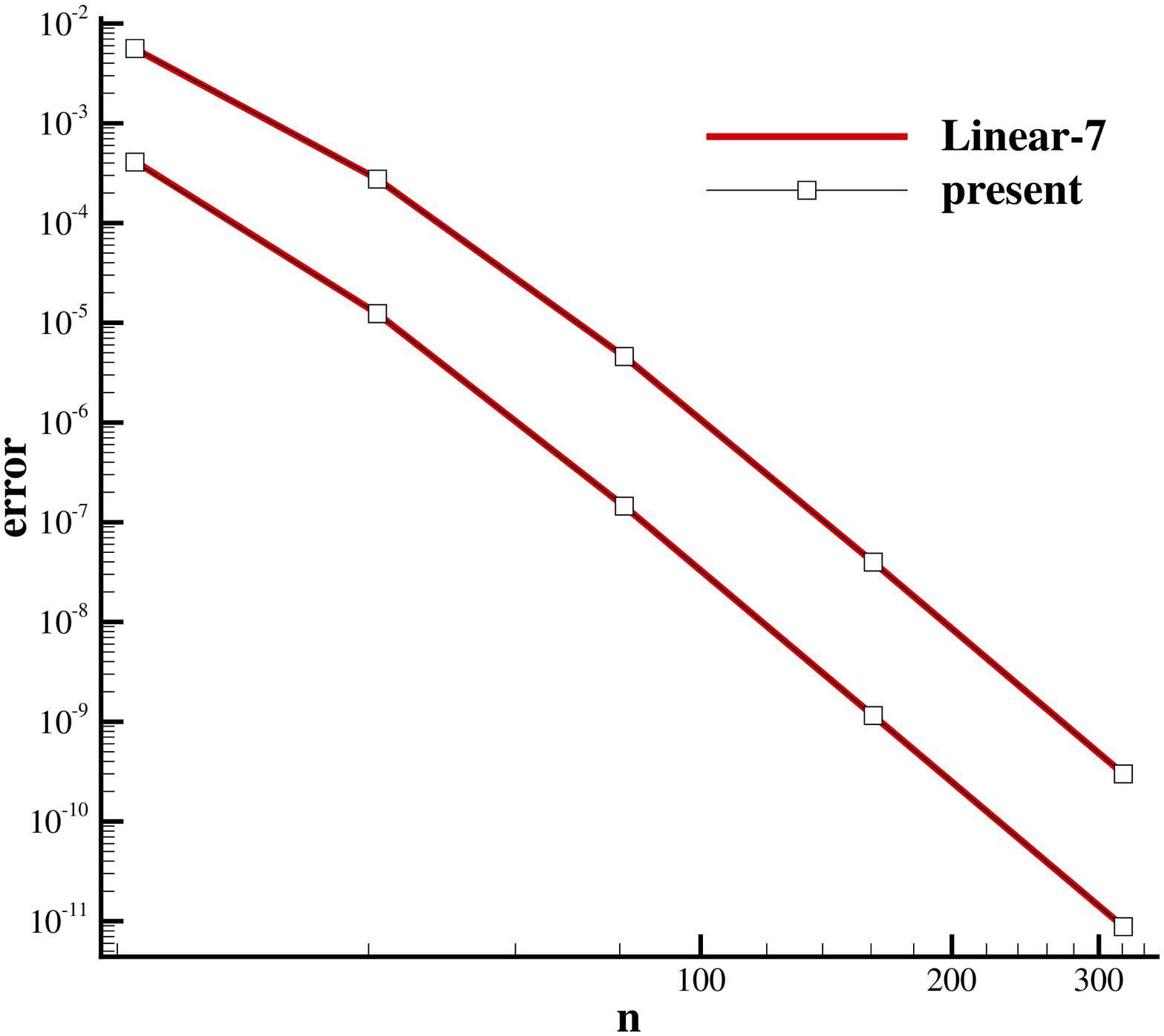}}}
\caption{\label{fig:f:burgers} The numerical results of the one-dimensional   Burgers equations: $t=0.5/\pi$. }
\end{center}
\end{figure}

\begin{figure}
 \centering
 \includegraphics[width=5.5cm]{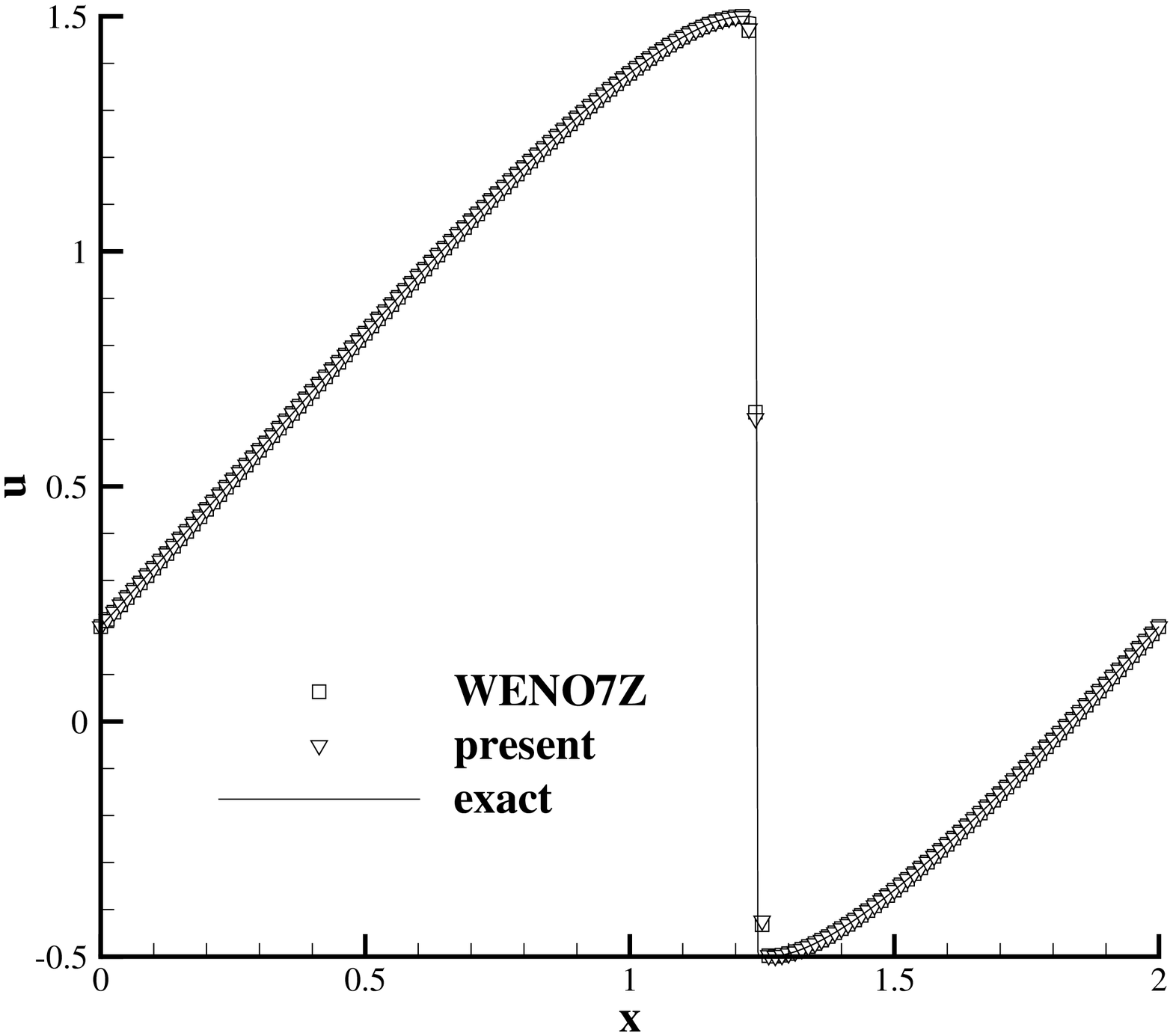}
 \caption{ \label{fig:f:burgers2}
  The numerical results of the one-dimensional   Burgers equations: $t=1.5/\pi$ and 160 grid points.}
\end{figure}

The solution is smooth if $0 \le t< 1/{\pi}$, and a discontinuity forms and starts to interact with the expansion wave if  $ t \ge 1/{\pi}$. Therefore,  the results at $t=0.5/\pi$ and $t=1.5/\pi$ are both given to show the continuous and discontinuous distributions.
In Fig.\ref{fig:f:burgers}, it can be found that the linear scheme is fully recovered by the presented method. The discontinuous results in Fig.\ref{fig:f:burgers2} proves that the presented method is oscillation-free, and its resolution is also comparable comparing with the seventh-order WENO-Z scheme.

\subsection{One dimensional Euler equations}
 Compressible  Euler equations are applied as physical models for the numerical test cases in this subsection. The  equation of state
 $p=(\gamma-1)\rho e$ with $\gamma=1.4$ is applied to close the Euler equations. The Roe average \cite{Roe1981} is exploited
for characteristic decomposition at the half-points, and the Global Lax-Friedrichs scheme is utilized for flux splitting. For the following one-dimensional cases uniformly distributed grid points are used. The third-order strongly stable Runge-Kutta scheme \cite{Gottlieb2001} with $CFL=0.6$
is used for time integration. The fifth-order TENO scheme, the seventh-order WENO-Z scheme \cite{Don2013}  and the presented method are applied for comparison.
\subsubsection{Isolated wave structure}

\begin{figure}[h!t]
\begin{center}
\subfigure[\label{fig:f:sod1} ]{
\resizebox*{7cm}{!}{\includegraphics{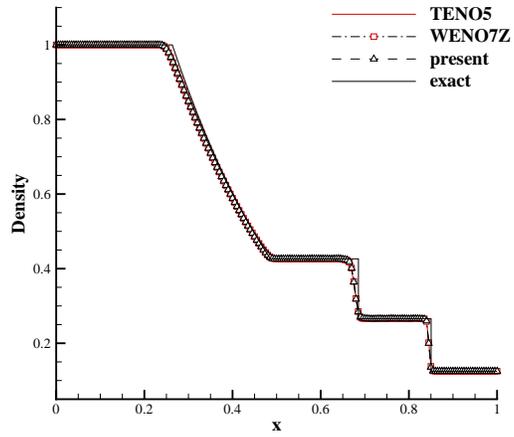}}}
\subfigure[\label{fig:f:sod2} ]{
\resizebox*{7cm}{!}{\includegraphics{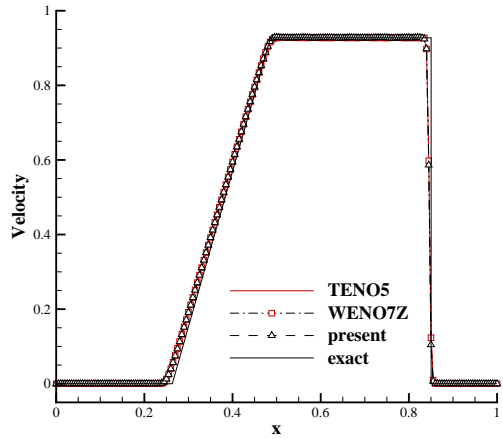}}}
\caption{\label{fig:f:sod} The results of Sod shock-tube simulation. }
\end{center}
\end{figure}

\begin{figure}[h!t]
\begin{center}
\subfigure[\label{fig:f:lax1} ]{
\resizebox*{7cm}{!}{\includegraphics{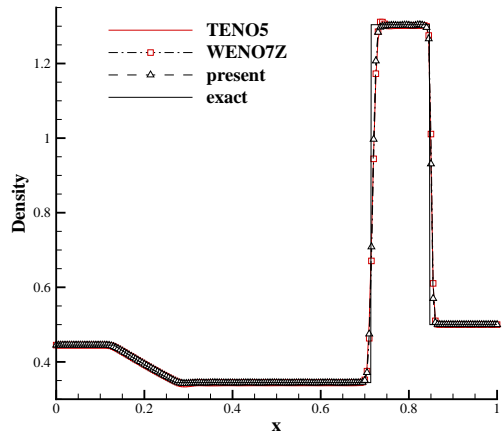}}}
\subfigure[\label{fig:f:lax2} ]{
\resizebox*{7cm}{!}{\includegraphics{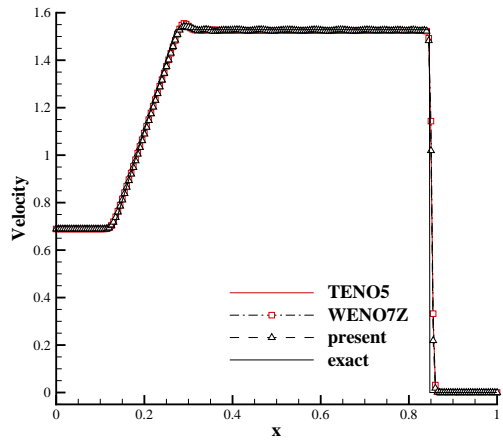}}}
\caption{\label{fig:f:lax}The results of Lax shock-tube simulation. }
\end{center}
\end{figure}

One dimensional shock-tube problems of Sod \cite{Sod1978} and Lax \cite{Lax1954} are used to evaluate the shock-capturing capability of the proposed scheme. The grid of $N=200$ points is used for spatial discretization.
The initial conditions for the Sod problem is
\begin{equation}
(\rho, u, p)=
\begin{cases}
\begin{matrix}
 (1,0,1)        &  \quad  x\in [0,0.5], \\
 (0.125,0,0.1)  & \quad  x\in  \left(0.5,1\right].
\end{matrix}
\end{cases}
\end{equation}
The solutions calculated using  at $t=0.2$ are given in Fig.\ref{fig:f:sod}. The ininital condition of Lax problem is
\begin{equation}
(\rho, u, p)=
\begin{cases}
\begin{matrix}
 (0.445, 0.698, 3.528)        &  \quad  x\in [0,0.5], \\
 (0.5,0,0.571)  & \quad  x\in  \left(0.5,1\right],
\end{matrix}
\end{cases}
\end{equation}
and the solutions at $t=0.14$ are given in Fig.\ref{fig:f:lax}.

The difference between the results of different methods is relatively small. However, the key here is that the presented method shows oscillation free results in capturing discontinuities. At the meantime, the seventh-order WENO-Z scheme shows a small overshoot across the contact discontinuity in Lax shock-tube simulation. Although the presented method  uses a simpler formula to achieve seventh-order interpolation, it  has captured a sharp and oscillation-free profile in the small situation.

\subsubsection{Shock wave-entropy wave interaction}

In order to investigate the performance in resolving smooth critical points, the one-dimensional test cases of Shu and Osher \cite{Shu1989} and  Titarev and Toro \cite{Titarev2004} are applied.

\begin{figure}[h!t]
\begin{center}
\subfigure[\label{fig:f:rho1} ]{
\resizebox*{7cm}{!}{\includegraphics{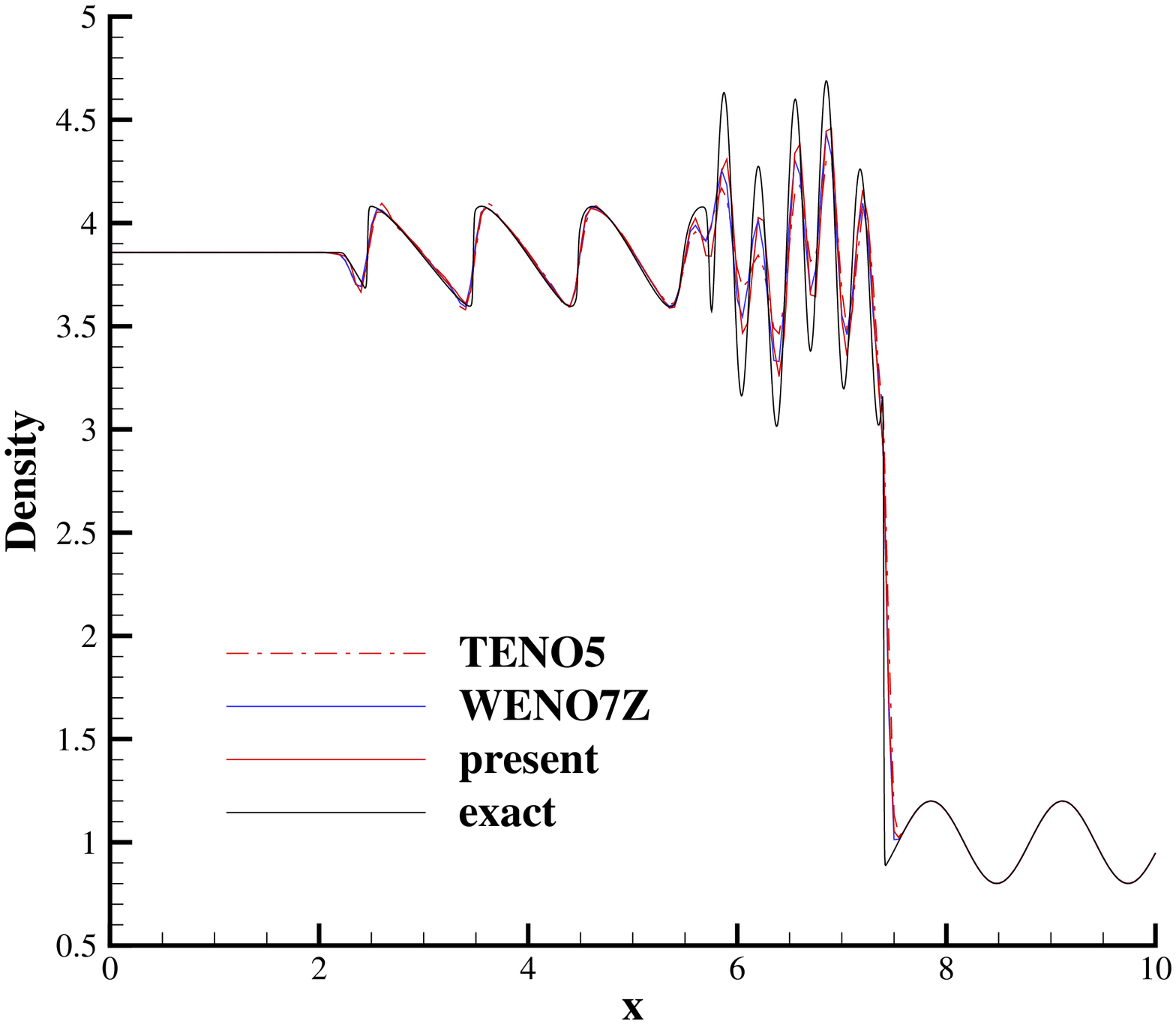}}}
\subfigure[\label{fig:f:rho2} ]{
\resizebox*{7cm}{!}{\includegraphics{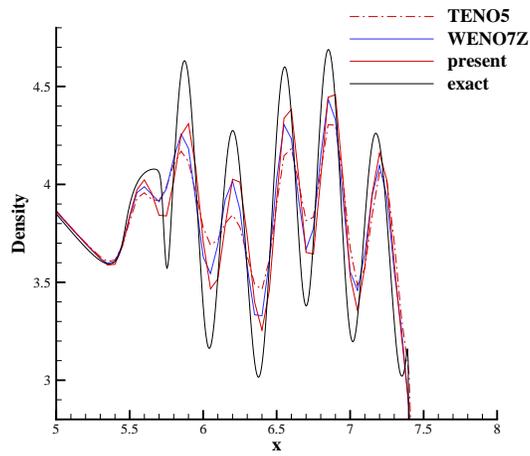}}}
\caption{\label{fig:f:rho} The density distribution of Shu-Osher shock-entropy wave interaction problem.}
\end{center}
\end{figure}

The shu and osher's case is discretized by 200 grid points, and the computation domain is $[0,10]$. The initial condition is designed as
\begin{equation} \label{eq:case}
(\rho, u, p)=
\begin{cases}
\begin{matrix}
(3.8571,2.6294,10.3333), & \text{if} \quad x \le 1, \\
(1+0.2\text{sin}(5x),0,1), & \text{if} \quad x > 1,
\end{matrix}
\end{cases}
\end{equation}
\noindent which consists of a right moving shock wave of Mach number 3 interacting with a  density
perturbation. The results at  $t=1.8$ are shown in Fig.\ref{fig:f:rho}, where the reference result is calculated by using  the five-point WENO-JS scheme on 2000 grid points, since there is not a theoretically \textit{exact} solution.

\begin{figure}[h!t]
\begin{center}
\subfigure[\label{fig:f:TT1} ]{
\resizebox*{7cm}{!}{\includegraphics{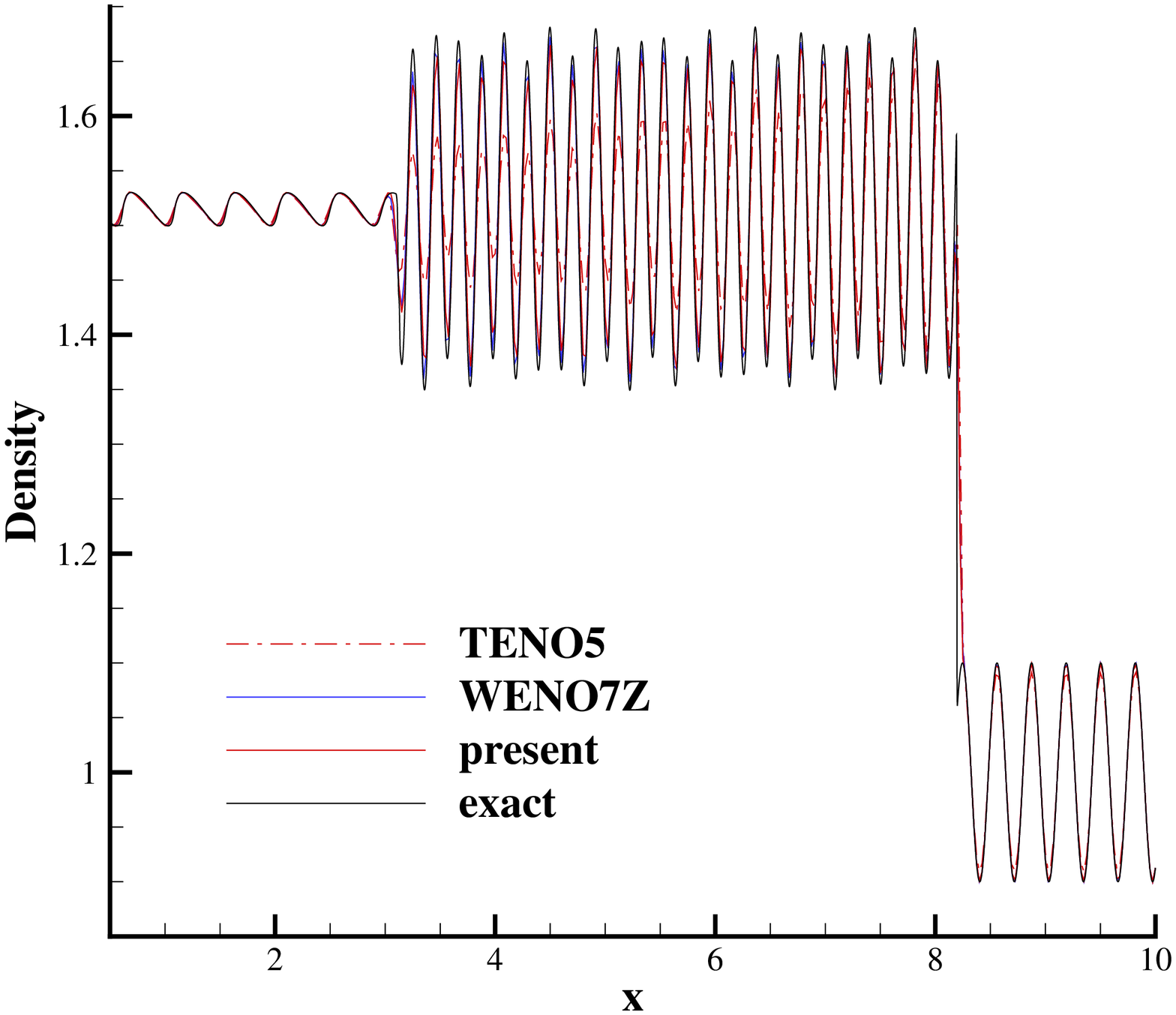}}}
\subfigure[\label{fig:f:TT2} ]{
\resizebox*{7cm}{!}{\includegraphics{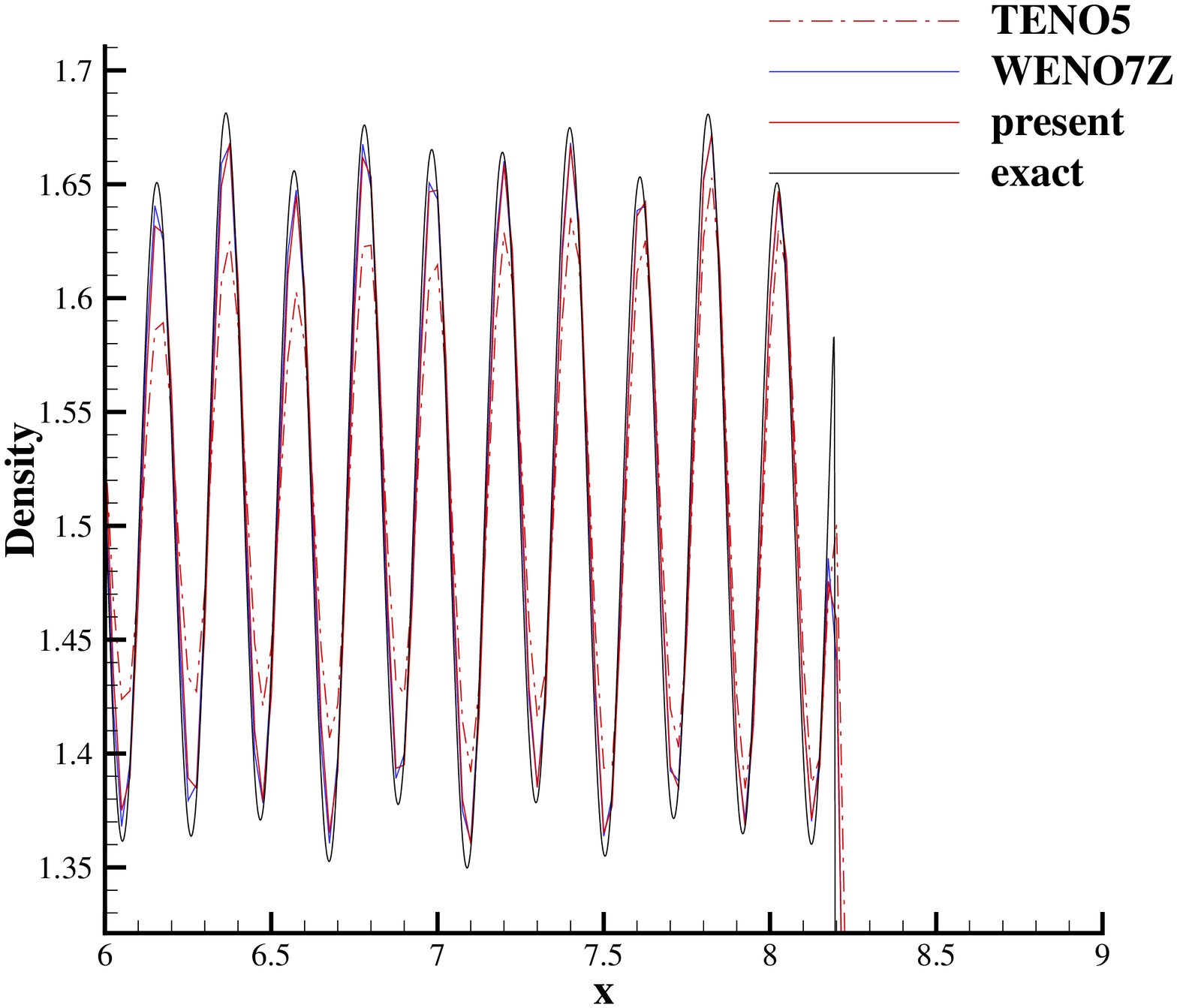}}}
\caption{\label{fig:f:TT} The density distribution of Titarev-Toro shock-entropy wave interaction problem. }
\end{center}
\end{figure}

The similar Titarev and Toro's case is discretized by 400 grid points, and the computation domain is $[0,10]$ as well.  The initial condition is designed as
\begin{equation} \label{eq:c_se}
(\rho, u, p)=
\begin{cases}
\begin{matrix}
(1.515695,0.523346,1.80500), & \text{if} \quad x \le 0.5, \\
(1+0.1\text{sin}(20\pi x),0,1), & \text{if} \quad x > 0.5,
\end{matrix}
\end{cases}
\end{equation}
\noindent which consists of a right moving shock wave   interacting with a high-frequency density
perturbation. The results at  $t=5$ are shown in Fig.\ref{fig:f:TT}, where the reference result is calculated by using  the five-point WENO-JS scheme on 8000 grid points, since there is not a theoretically \textit{exact} solution.

 It can be found that significant advantage in achieving the reference solution has been found by using the presented method, comparing with the fifth-order TENO scheme, in the both two cases. At the meantime, comparing with the seventh-order WENO-Z scheme, the presented method still shows advantage  in shu and osher's case, and similar resolution is shown in  Titarev and Toro's case.
 Therefore, although a simpler formula has been applied in the presented method, we can see that the scheme is still capable to provide high resolution and oscillation-free results.

\section{Concluding remarks} \label{sec:Conclusions}

In this article, a simple extending strategy  based on the TENO framework is introduced. Specifically, the smoothness measurement of the fifth-order TENO scheme is used, except that the small parameter $\epsilon$ is redefined. Therefore, without high computation cost or complex procedure, the presented method achieve up to seventh-order of accuracy by simply exploiting the neighboring information, if the smooth field is sufficiently large.

In the presented method, the smoothness information is no longer restricted within a five-point full stencil, as in the typical five-point TENO or WENO scheme. Instead, all the smoothness information is stored in spatial interpolation step which can be  using various high-order polynomials of different  order of accuracy.  This flexibility is gained by exploring the ENO-like stencil selection procedure of TENO schemes.

Since higher-order interpolation can be achieved, and the spatial reconstruction crossing discontinuity is completely avoided, the presented method shows satisfactory performance in various numerical cases, with or without discontinuity. Especially, although the formula is as simple as the fifth-order TENO scheme, the performance is comparable or even better than the seventh-order WENO-Z scheme.
 Moreover, if there are more smooth stencils being continuously located, higher order polynomial is a potential choice to further improve the accuracy.

\section*{Acknowledgments}
The first author is supported by the Open fund from State Key Laboratory of Aerodynamics (No. SKLA20180302), the second author is supported by  the National Natural Science Foundation of China (No. 11872144) and the third author is supported by the grant   20187413071020008.

\bibliography{ref}
\end{document}